\documentclass[twocolumn,conference]{IEEEtran}
\usepackage[T1]{fontenc}
\usepackage[latin9]{inputenc}
\usepackage{float}
\usepackage{units}
\usepackage{textcomp}
\usepackage{amsmath}
\usepackage{graphicx}
\usepackage[unicode=true,
 bookmarks=true,bookmarksnumbered=true,bookmarksopen=true,bookmarksopenlevel=1,
 breaklinks=false,pdfborder={0 0 0},pdfborderstyle={},backref=false,colorlinks=false]
 {hyperref}
\hypersetup{pdftitle={Your Title},
 pdfauthor={Your Name},
 pdfpagelayout=OneColumn, pdfnewwindow=true, pdfstartview=XYZ, plainpages=false}

\makeatletter

\providecommand{\tabularnewline}{\\}
\floatstyle{ruled}
\newfloat{algorithm}{tbp}{loa}
\providecommand{\algorithmname}{Algorithm}
\floatname{algorithm}{\protect\algorithmname}

\usepackage[caption=false,font=footnotesize]{subfig}
\usepackage{cite}
\renewcommand{\baselinestretch}{0.935}

\makeatother

\begin{document}
\title{A Graphical Correlation-Based Method for Counting the Number of Global
8-Cycles on the SCRAM Three-Layer Tanner Graph}
\author{\IEEEauthorblockN{Sally Nafie, Joerg Robert, Albert Heuberger}\IEEEauthorblockA{Lehrstuhl für Informationstechnik mit dem Schwerpunkt Kommunikationselektronik
(LIKE)\\
Friedrich-Alexander Universität Erlangen-Nürnberg (FAU), 91058 Erlangen,
Germany\\
\{sally.nafie, joerg.robert, albert.heuberger\}@fau.de}%
\noindent\begin{minipage}[t]{1\textwidth}%
This work has been submitted to the IEEE for possible publication.
Copyright may be transferred without notice, after which this version
may no longer be accessible.%
\end{minipage}}
\maketitle
\begin{abstract}
This paper presents a novel graphical approach that counts the number
of global 8-cycles on the SCRAM three-layer Tanner graph. SCRAM, which
stands for Slotted Coded Random Access Multiplexing, is a joint decoder
that is meets challenging requirements of 6G. At the transmitter side,
the data of the accommodated users is encoded by Low Density Parity
Check (LDPC) codes, and the codewords are transmitted over the shared
channel by means of Slotted ALOHA. Unlike the state-of-the-art sequential
decoders, the SCRAM decoder jointly resolves collisions and decodes
the LDPC codewords, in a similar analogy to Belief Propagation on
a three-layer Tanner graph. By leveraging the analogy between the
two-layer Tanner graph of conventional LDPC codes and the three-layer
Tanner graph of SCRAM, the well-developed analysis tools of classical
LDPC codes could be utilized to enhance the performance of SCRAM.
In essence, the contribution of this paper is three-fold; First it
proposes the methodology to utilize these tools to assess the performance
of SCRAM. Second, it derives a lower bound on the shortest cycle length
of an arbitrary SCRAM Tanner graph. Finally, the paper presents a
novel graphical method that counts the number of cycles of length
that corresponds to the girth.
\end{abstract}


\section{Introduction }

The emerging 6G technology is provisioned to provide a wide spectrum
of services and use cases such as Further enhanced Mobile Broadband
(FeMBB), Extremely-Reliable $\textrm{Low}\textrm{-}\textrm{Latency}\:\textrm{Communication\ensuremath{\:\textrm{(ERLLC)}}}$,
$\textrm{ultra-massive}\;\textrm{Machine}$ Type Communications (umMTC),
Long-Distance High-Mobility Communications (LDHM), and Extremely-Low-Power
Communications (ELPC) \cite{miraz2015review}, in addition to numerous
services that fall under the umbrella of Internet of Everything (IoE)
\cite{saad2019vision}. The diverse services and use cases, expected
to be supported by 6G impose challenges on the wireless systems, in
terms of massive connectivity, reliability, latency, and complexity
\cite{chowdhury20206g}.

The state-of-the-art techniques resort to Non-Orthogonal Multiple
Access (NOMA) \cite{ding2015cooperative,islam2016power,reddy2021analytical}
as a leading approach adopted in 5G to support overloading. NOMA schemes
are classified into Power $\textrm{Domain}\:\textrm{(PD)}$ and Code
Domain (CD) NOMA \cite{srivastava2021non}. In PD-NOMA \cite{maraqa2020survey,lee2020user,gupta2020user},
the system exploits the power-domain as the key ingredient that differentiates
the users. Alternatively, NOMA schemes could be designed in the code
domain, where each user is assigned different code \cite{jehan2022comparative,liu2021sparse}.
In the literature, two viable CD-NOMA schemes are presented; Pattern
Division Multiple Access (PDMA) \cite{chen2016pattern} and Sparse
Code Multiple Access (SCMA) \cite{nikopour2013sparse}.

In order to support overloading, NOMA schemes rely on thoroughly coordinating
the resource allocation among the users. Consequently, these techniques
perform poorly in terms of scalability. To ensure reliability, most
of the proposed schemes adopt repetition coding \cite{al2014uplink},
or a concatenation of repetition codes with more sophisticated Forward
Error Correction (FEC) codes \cite{chaturvedi2022tutorial}. According
to coding theory, repetition coding does not make the best use of
the resources. To meet the low-latency constraint, some CD-NOMA schemes
suggest a grant-free variant of their proposed scheme \cite{zhang2022hybrid}.
However, because these schemes strongly rely on coordinating the collisions,
a grant-free scheme thereof is provisioned to deteriorate the performance.
The techniques \cite{gollakota2008zigzag,tehrani2011sigsag} transmit
replicas of the encoded packets by means of Random Access. These techniques
not only lose the degrees of freedom because of the repetition coding,
but also do not support overloading as the number of repetitions corresponds
to the number of adopted users. 

At the receiver side, the state-of-the-art NOMA schemes adopt a sequential
decoding structure, portrayed in a Multiuser Detector (MUD) block
that resolves the collisions, followed by a bank of FEC decoders to
decode the adopted FEC code \cite{xiao2019low,herath2020low,iswarya2021survey,ling2017multiple}.
Some schemes tend to further optimize their decoding performance by
adopting Turbo-like decoders which are also known as Iterative Detection
and Decoding (IDD) \cite{ren2016advanced}. Such decoding structure
suggests extending a feedback link between the output of the the bank
of FEC decoders and the input of the MUD detector. In essence, IDD
could be regarded as an iterative sequential decoder.

In \cite{nafie2018scram}, a novel approach dubbed SCRAM is proposed.
SCRAM, which stands for Slotted Coded Random Access Multiplexing,
is a hybrid joint decoder, that combines the ultra-low latency privilege
of Slotted ALOHA (SA) \cite{munari2015multi}, with the robust performance
of Low Density Parity Check (LDPC) codes \cite{gallager1962low}.
At the transmitter side, the data of the various users are encoded
by means of LDPC codes, and the codewords are transmitted over the
shared channel by means of SA. The uncoordinated access results in
collisions among the transmitted packets. Unlike the sequential decoders
presented in the literature, the essence of the hybrid SCRAM decoder
lies in its ability to jointly resolve the collisions, and decode
LDPC, in a similar fashion to Belief Propagation (BP) \cite{mackay1997near}
on a joint three-layer Tanner graph.

Due to the analogy between the two-layer Tanner graph \cite{tanner1981recursive}
representation of LDPC codes, and the three-layer Tanner graph of
the proposed SCRAM, conventional methods that analyze the performance
of classical LDPC codes, could be adopted to analyze and further optimize
the performance of the proposed SCRAM. In classical LDPC codes, a
visible error floor in the PER performance at high SNR region is caused
by the presence of cycles in the adjacent Tanner graph \cite{karimi2012message}.

Numerous techniques are proposed in the literature to assess the cycle
profile of classical LDPC codes. By mapping the three-layer Tanner
graph of SCRAM, these techniques could be adopted to analyze the cycle
profile SCRAM. However, the complexity of these techniques, when adopted
for SCRAM, grow exponentially in both the number of accommodated users,
and the deployed LDPC codeword lengths. For that reason, this paper
presents a thorough analysis tailored to the SCRAM three-layer Tanner
graph, in addition to a derivation of the shortest cycle length, known
as the girth. In essence, the detailed analysis is utilized to design
a graphical approach that quantifies the number of cycles of girth
length, on any arbitrary SCRAM Tanner graph.

The rest of this paper is organized as follows. $\textrm{Section}\:$II
covers the system model of SCRAM. In Section III, the utilization
of the cycle counting algorithms of calssical LDPC codes, in conjunction
with SCRAM is tackled. In Section IV, a detailed analysis of the SCRAM
three-layer Tanner graph is presented, in addition to the derivation
of the SCRAM girth. Section V proposes a novel graphical algorithm
that quantifies the number of cycles whose length equals to the girth.
Moreover, Section VI shows the results of the proposed graphical method.
Finally, Section VII summarizes the paper and sheds light on the potential
future implications.

\section{SCRAM Preliminaries}

\subsection{System Model}

The Hybrid SCRAM system incorporates $N_{u}$ users sharing the SA
wireless medium. Each user $U_{n_{u}},\;\text{\ensuremath{\forall}}n_{u}=1,\ldots,N_{u}$,
has an information packet, $\mathbf{b^{\left(\mathit{n_{u}}\right)}}$,
of length $k_{n_{u}}$ bits. This packet is encoded with an LDPC encoder,
of code rate $r_{n_{u}}=k_{n_{u}}/n_{n_{u}}$, producing an output
codeword, $\mathbf{c^{\left(\mathit{n_{u}}\right)}}$, of length $n_{n_{u}}$
bits. Without loss of generality, the encoded bits are mapped to BPSK
modulated symbols. The vector of modulated symbols of user $U_{n_{u}}$,
is given by $\mathbf{x^{\left(\mathit{n_{u}}\right)}}$. The BPSK
symbols are then to be transmitted using OFDM. Incorporating OFDM
is twofold: to provide slot synchronization due to the gridded analogy
between SA and OFDM subcarriers, and to translate the multipath fading
channel to multiple flat fading subchannels. Prior to transmission,
each user, $U_{n_{u}}$, randomly and independently chooses $n_{n_{u}}$
SA slots (OFDM subcarriers) to transmit its modulated encoded codeword.
Let $N_{s}$ denote the total number of available slots per SA frame.
In this case, the channel load -- defined as the number of useful
information bits per SA slot -- is given by $D=\sum_{n_{u}=1}^{N_{u}}k_{n_{u}}/N_{s}$.

At the receiver side, a joint decoding of the contended LDPC codewords
is performed iteratively. Unlike$\:\textrm{the}\;\textrm{sequential}$
decoders presented in the literature, the SCRAM mechanism incorporates
a joint three-layer Tanner graph that allows for the joint decoding.
The idea is inspired by BP decoding of graph-based codes such as LDPC
codes. 

\subsection{Three-Layer Tanner Graph Representation}

Assuming a SCRAM system that accommodates $N_{u}$ users, the corresponding
joint three-layer Tanner graph comprises a set of variable nodes representing
the transmitted symbols from all the users. These variable nodes are
bounded by two layers of check nodes. The first check nodes layer
corresponds to the slots of the shared SA medium. The other layer
is driven from the conventional parity check nodes in the BP decoding
of LDPC codes. Let $N_{v}$ denote the number of variable nodes. Each
transmitted symbol is represented by a variable node. Thus, $N_{v}=\sum_{n_{u}=1}^{^{N_{u}}}n_{n_{u}}$,
where $n_{n_{u}}$ represents the number of transmitted symbols of
user $U_{n_{u}}$ per SA frame. The number of SA check nodes, denoted
by $N_{s}$, is allocated according to the available resources. The
total number of LDPC check nodes is given by $N_{l}=\sum_{n_{u}=1}^{^{N_{u}}}m_{n_{u}}$,
where $m_{n_{u}}\geq n_{n_{u}}-k_{n_{u}}$, denotes the number of
LDPC parity check equations of user $U_{n_{u}}$.

The edges connecting the variable nodes and the SA check nodes depend
on the random selection of the transmission slots of each user. This
means that a contended SA check node can be connected to more than
one variable node. Meanwhile, the connections of the variable nodes
to the LDPC check nodes are fully determined by the deployed LDPC
encoder at each user's transmit terminal.

\begin{figure}[t]
\begin{centering}
\includegraphics[width=1\columnwidth]{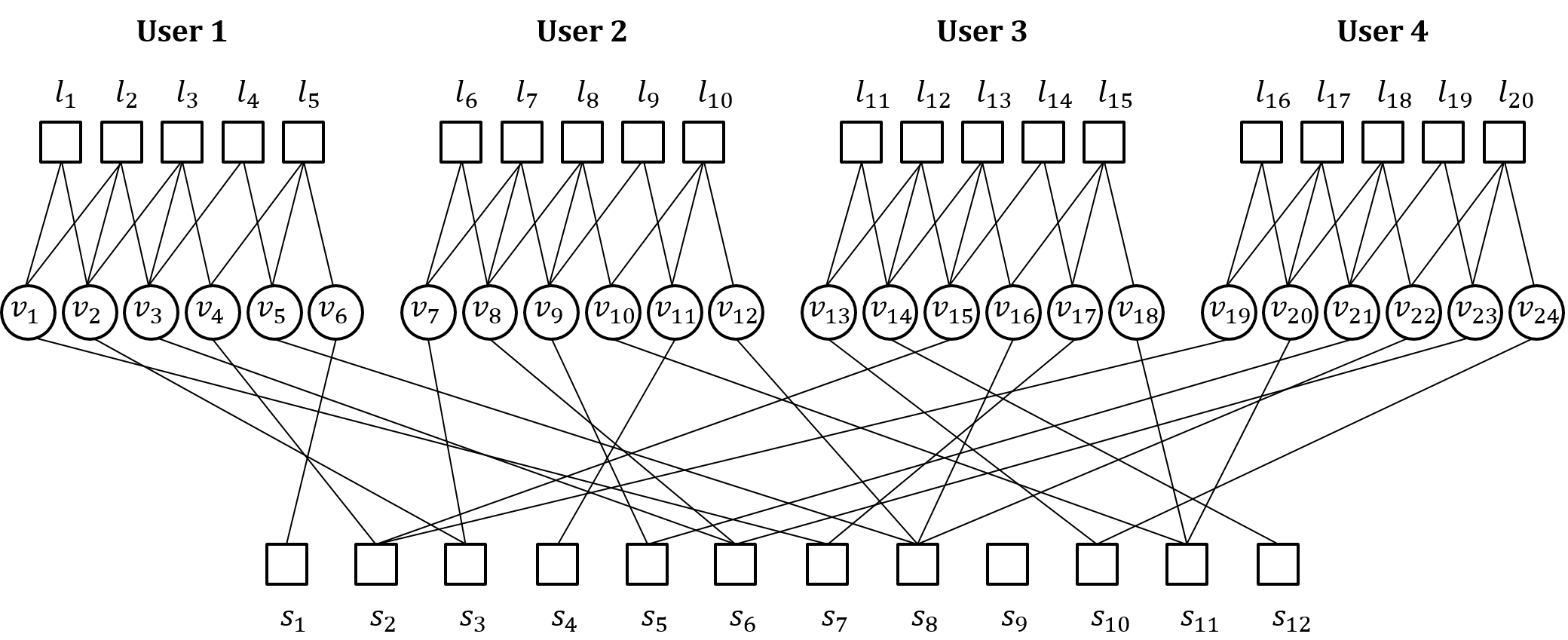}
\par\end{centering}
\caption{Three-Layer Tanner graph example of SCRAM system with $N_{u}=4$ users,
each tranmitting $n_{n_{u}}=6$ LDPC encoded symbols over a system
with $N_{s}=12$ SA slots}
\label{Tanner Example}

\end{figure}

For illustration, consider the three-layer Tanner graph of the SCRAM
model shown in Figure \ref{Tanner Example}. The model incorporates
$N_{u}=4$ users, each transmitting $n_{n_{u}}=6$ modulated symbols.
It is assumed that the four users adopt identical LDPC encoders with
$n_{n_{u}}=6$ variable nodes, and $m_{n_{u}}=5$ LDPC check nodes,
per user. This means that the three-layer Tanner graph of such a model
consists of $N_{v}=24$ variable nodes, $N_{l}=20$ LDPC check nodes.
Moreover, the graph has a set of $N_{s}=12$ SA check nodes, that
correspond to 12 allocated frequency subcarriers. It is also assumed
that each user blindly selects six subcarriers to transmit its six
modulated symbols.

\subsection{Iterative Joint SCRAM Decoding}

Algorithm \ref{Alg. SCRAM} depicts the steps of the joint SCRAM scheme
for one decoding iteration. The variable nodes concurrently communicate
with both the SA and the LDPC check nodes, on the three-layer Tanner
graph that comprises $N_{v}$ variable nodes, $N_{s}$ SA check nodes,
and $N_{l}$ LDPC check nodes.

\subsubsection{Check Nodes to Variable Nodes}

In the first half of the iteration, the SA and LDPC check nodes calculate
their provisioned Log Likelihood Ratios (LLRs), and send them to their
corresponding variable nodes. 

\paragraph{SA Check Nodes to Variable Nodes}

The SA check node update is illustrated in steps 1 through 13, such
that $S_{n_{s},n_{v}}$ represents the LLR transmitted from SA check
node $s_{n_{s}}$ to variable node $v_{n_{v}}$, if $v_{n_{v}}$ belongs
to the set $A_{s_{n_{s}}}$, of connected variable nodes to SA check
node $s_{n_{s}}$. Assuming $V_{n_{s},\acute{n_{v}}}^{\left(S\right)}$
denotes the LLR that SA check node $s_{n_{s}}$ received from its
corresponding variable node $v_{\acute{n_{v}}}$in the previous iteration,
in step 10, $s_{n_{s}}$, calculates the conditional probability of
the received signal given every possible combination of its collided
symbols. In steps 5 through 9, SA check node $s_{n_{s}}$, computes
the joint apriori probability of these vector combinations, while
steps 3 and 4 calculate the probability of the individual elements
within these vectors.

\noindent The following notations are adopted. For $n_{s}=1,\cdots,N_{s}$,
$y_{n_{s}}$ represents the channel received signal at SA slot $s_{n_{s}}$.
Moreover, $M_{n_{s}}^{\left(\pm\right)}$ denotes the cardinality
of the set of possible vectors of all the collided symbols at SA check
node $s_{n_{s}}$, that assume that the symbol represented by variable
node $v_{n_{v}}$ is $+1$ and $-1$, respectively. For $\acute{m}=1,\cdots,M_{n_{s}}^{\left(\pm\right)}$,
let $\mathbf{\mathbf{\check{x}_{\mathit{\acute{m}}}^{\left(\mathit{n_{v},\mathrm{+1}}\right)}}}$
and $\mathbf{\mathbf{\check{x}_{\mathit{\acute{m}}}^{\left(\mathit{n_{v},\mathrm{-1}}\right)}}}$
be a vector that corresponds to one of the $M_{n_{s}}^{\left(\pm\right)}$
combination vectors, that assume that the symbol represented by $v_{n_{v}}$
is $+1$ and $-1$, respectively. Furthermore, $\mathrm{P}(\mathbf{\mathbf{\check{x}_{\acute{\mathit{m}}}^{\left(\mathit{n_{v},\mathrm{+1}}\right)}}})$
and $\mathrm{P}(\mathbf{\mathbf{\check{x}_{\acute{\mathit{m}}}^{\left(\mathit{n_{v},\mathrm{-1}}\right)}}})$
denote the apriori probability of the vectors $\mathbf{\mathbf{\check{x}_{\mathit{\acute{m}}}^{\left(\mathit{n_{v},\textrm{+1}}\right)}}}$
and $\mathbf{\mathbf{\check{x}_{\mathit{\acute{m}}}^{\left(\mathit{n_{v},\textrm{-1}}\right)}}}$,
respectively. In addition, $\check{x}_{\acute{m},\acute{d}}^{\left(n_{v},+1\right)}$and
$\check{x}_{\acute{m},\acute{d}}^{\left(n_{v},-1\right)}$ represent
a possible value for the modulated symbol represented by the $\acute{d}^{\textrm{th}}$
variable node that collides at SA check node $s_{n_{s}}$, in the
vector $\check{\mathbf{x}}_{\acute{\mathit{m}}}^{\left(\mathit{n_{v}}\mathrm{,+1}\right)}$and
$\mathbf{\check{x}_{\mathit{\acute{m}}}^{\left(\mathit{n_{v}}\mathrm{,-1}\right)}}$,
respectively. Moreover, $h_{s_{n_{s}},\acute{d}}$ denotes the estimated
fading coefficient of the channel between SA check node $s_{n_{s}}$
and the $\acute{d}^{\textrm{th}}$ variable node that collides at
it. Finally, $\sigma^{2}$ is the noise variance of the complex AWGN
channel.

\paragraph{LDPC Check Nodes to Variable Nodes}

Inspired by BP decoding of classical LDPC codes, Steps 14 through
18 give the LDPC check node update, such that $L_{n_{l},n_{v}}$,
represents the LLR to be transmitted from LDPC check node $l_{n_{l}}$,
to variable node $v_{n_{v}}$, while $V_{n_{l},\acute{n_{v}}}^{\left(L\right)}$
represents the LLR received by LDPC check node $l_{n_{l}}$ from variable
node $v_{\acute{n_{v}}}$, in the previous iteration. Moreover, $A_{l_{n_{l}}}$
denotes the set of variable nodes connected to LDPC check node $l_{n_{l}}$. 

\subsubsection{Variable Nodes to Check Nodes}

Upon receiving the LLRs from the SA and LDPC check nodes, the variable
nodes calculate new LLRs to be transmitted to their corresponding
check nodes, both SA and LDPC, in the next half of the iteration.
Steps 19 through 24 highlight the variable node update, such that
$V_{n_{l},n_{v}}^{\left(L\right)}$ and $V_{n_{s},n_{v}}^{\left(S\right)}$
represent the information transmitted from variable node, $v_{n_{v}}$
to LDPC check node $l_{n_{l}}$, and SA check node $s_{n_{s}}$, respectively.
Moreover, $L_{\acute{n_{l}},n_{v}}$ and $S_{\acute{n_{s}},n_{v}}$
represent the LLR, that variable node, $v_{n_{v}}$, has received
in the first half of the iteration from LDPC check node, $l_{\acute{n_{l}}}$,
and SA check node, $s_{\acute{n_{s}}}$, respectively. Furthermore,
$A_{v_{n_{v}}}^{\left(L\right)}$ and $A_{v_{n_{v}}}^{\left(S\right)}$,
respectively denote the set of LDPC check nodes, and the set of SA
check nodes, connected to variable node $v_{n_{v}}$.

\begin{algorithm}[tbh]
\caption{Joint SCRAM Decoding}
\label{Alg. SCRAM}

//SA Check Node Update

1:\textbf{ for} $n_{s}=1:N_{s}$ \textbf{do}

2:$\;\;\;\;$\textbf{for} $n_{v}\in A_{s_{n_{s}}}$\textbf{ do}

3:$\;\;\;\;\;\;\;\;\mathrm{P}\left(\check{x}_{\acute{m},\acute{d}}^{\left(n_{v},+1\right)}\right)=\frac{1}{1+\exp\left[-V_{n_{s},\acute{n_{v}}}^{\left(S\right)}\right]}$

4:$\;\;\;\;\;\;\;\;\mathrm{P}\left(\check{x}_{\acute{m},\acute{d}}^{\left(n_{v},-1\right)}\right)=\frac{-V_{n_{s},\acute{n_{v}}}}{1+\exp\left[-V_{n_{s},\acute{n_{v}}}^{\left(S\right)}\right]}$

5:$\;\;\;\;\;\;\;\;$\textbf{if} $iter\equiv1$ \textbf{then}

6:$\;\;\;\;\;\;\;\;\;\;\;\;\mathrm{P}(\mathbf{\check{x}_{\acute{\mathit{m}}}^{\left(\mathit{n_{v}}\mathrm{,\pm1}\right)}})=\frac{1}{M_{n_{s}}^{\left(\pm\right)}}$

7:$\;\;\;\;\;\;\;\;$\textbf{else}

8:$\mathrm{\;\;\;\;\;\;\;\;\;\;\;\;P}(\mathbf{\check{x}_{\mathit{\acute{m}}}^{\left(\mathit{n_{v}}\mathrm{,\pm1}\right)}})=\ensuremath{\prod_{\substack{\acute{d}=1\\
v_{n_{v}}\neq v_{s_{n_{s}},\acute{d}}
}
}^{d_{s_{n_{s}}}}\mathrm{P}\left(\check{x}_{\acute{m},\acute{d}}^{\left(n_{v},\pm1\right)}\right)}$

9:$\;\;\;\;\;\;\;\;$\textbf{end if}

10:$\mathrm{\;\;\;\;\;\;\;P}\left(y=y_{n_{s}}|\mathbf{\check{x}_{\acute{\mathit{m}}}^{\left(\mathit{n_{v}}\mathrm{,\pm1}\right)}}\right)=$

$\;\;\;\;\;\;\;\;\;\;\;\;\;\;\;\frac{1}{\pi\sigma^{2}}\exp\left[\frac{1}{\sigma^{2}}\cdot\left|y_{n_{s}}-\ensuremath{\sum\limits _{\acute{d}=1}^{d_{s_{n_{s}}}}h_{s_{n_{s}},\acute{d}}\;}\check{x}_{\acute{m},\acute{d}}^{\left(n_{v},\pm1\right)}\right|^{2}\right]$

\smallskip{}

11:$\;\;\;\;\;\;\;S_{n_{s},n_{v}}=\ln\left[\frac{\ensuremath{\sum\limits _{\acute{m}=1}^{M_{n_{s}}^{\left(\pm\right)}}\left[\mathrm{P}(y=y_{n_{s}}|\check{\mathbf{x}}_{\acute{m}}^{\left(n_{v},+1\right)})\cdot\mathrm{P}\left(\check{\mathbf{x}}_{\acute{m}}^{\left(n_{v},+1\right)}\right)\right]}}{\sum\limits _{\acute{m}=1}^{M_{n_{s}}^{\left(\pm\right)}}\left[\mathrm{P}(y=y_{n_{s}}|\check{\mathbf{x}}_{\acute{m}}^{\left(n_{v},-1\right)})\cdot\mathrm{P}\left(\check{\mathbf{x}}_{\acute{m}}^{\left(n_{v},-1\right)}\right)\right]}\right]$

12:$\;\;\;$\textbf{end for}

13:\textbf{ end for}

//LDPC Check Node Update

14:\textbf{ for} $n_{l}=1:N_{l}$ \textbf{do}

15:$\;\;\;\;$\textbf{for} $n_{v}\in A_{l_{n_{l}}}$\textbf{ do}

16:$\;\;\;\;\;\;\;\;L_{n_{l},n_{v}}=-2\tanh^{-1}\left(\prod_{\substack{v_{\acute{n_{v}}}\epsilon\mathbf{v}_{l_{n_{l}}}\\
v_{\acute{n_{v}}}\neq v_{n_{v}}
}
}\tanh\left(-\frac{V_{n_{l},\acute{n_{v}}}^{\left(L\right)}}{2}\right)\right)$

17:$\;\;\;$\textbf{end for}

18:\textbf{ end for}

//Variable Node Update

19:\textbf{ for} $n_{v}=1:N_{v}$ \textbf{do}

20:$\;\;\;\;$\textbf{for} $n_{l}\in A_{v_{n_{v}}}^{\left(L\right)}$,$\;n_{s}\in A_{v_{n_{v}}}^{\left(S\right)}$\textbf{
do}

21:$\;\;\;\;\;\;\;\;V_{n_{l},n_{v}}^{\left(L\right)}=\ensuremath{\sum\limits _{s_{\acute{n_{s}}}\in\mathbf{\boldsymbol{s}}_{v_{n_{v}}}}S_{\acute{n_{s}},n_{v}}}+\sum_{\substack{l_{\acute{n_{l}}}\in\mathbf{\mathbf{l}}_{v_{n_{v}}}\\
l_{\acute{n_{l}}}\neq l_{n_{l}}
}
}L_{\acute{n_{l}},n_{v}}$

22:$\;\;\;\;\;\;\;\;V_{n_{s},n_{v}}^{\left(S\right)}=\sum_{\substack{s_{\acute{n_{s}}}\in\boldsymbol{\mathbf{s}}_{v_{n_{v}}}\\
s_{\acute{n_{s}}}\neq s_{n_{s}}
}
}S_{\acute{n_{s}},n_{v}}+\ensuremath{\sum\limits _{l_{\acute{n_{l}}}\in\mathbf{\mathbf{l}}_{v_{n_{v}}}}L_{\acute{n_{l}},n_{v}}}$

23:$\;\;\;$\textbf{end for}

24:\textbf{ end for}

\end{algorithm}

\section{Cycle-Profile of SCRAM}

On a Tanner graph, a cycle is defined as a path that starts and terminates
at the same node, and goes through a sequence of nodes via their connected
edges \cite{karimi2012message}. For the path to be called a cycle,
the edges, and all the nodes except the terminal ones have to be distinct.
If the first edge in the path is the same as the last edge, the path
is rather referred to as a closed walk, and is not considered a cycle
\cite{karimi2012message}. A cycle of length $L$ is simply referred
to as an $L$- cycle. The girth of a Tanner graph is denoted by $g$,
and is defined as the length of the shortest cycle within the graph
\cite{li2015improved}. A cycle profile is a quantitative analysis
of the number of short cycles within a Tanner graph. These short cycles
are namely those whose length is less than double the girth \cite{karimi2012message}.

\subsection{Cycle-Profile of Classical LDPC Codes}

An $(n,k)$ LDPC code is fully described by a Tanner graph that comprises
$n$ variable nodes, and $m\ge n-k$ check nodes \cite{johnson2006introducing}.
The presence of short cycles in the Tanner graph hinders the performance
of BP, and leads to error propagation \cite{xiao2019reed}. For that
purpose, many algorithms that quantify and evaluate the cycle profile
of a given LDPC code were proposed.

Two algorithms are chosen in the sequel for evaluating the cycle profile
of a given LDPC code, and later for extending the analysis to the
three-layer Tanner graph of SCRAM. The first algorithm \cite{karimi2012message}
adopts an iterative message-passing algorithm, that tracks for every
node within the Tanner graph, the number of iterations required by
a monomial variable to make a full-cycle back to this initiating node.
The second algorithm \cite{li2015improved} debates that a cycle that
starts and ends at one node, can be traced at an intermediate point
half-way within the cycle length. These two algorithms are going to
be referred to as Full-Cycle and Half-Cycle counting algorithms, respectively.

\subsubsection{Full-Cycle Algorithm}

To calculate the number of cycles of particular length that pass through
a specific node, the algorithm is initialized at $t=0$, by allowing
this node to transmit a different monomial (dummy variable of unit
power) over each of its connected edge, while all the other nodes
within the set would send unity. The message passing algorithm is
activated, where at every time instant, only one set of nodes is active
and is allowed to pass information to the corresponding set of nodes.
The information that a node passes on any of its connected edges corresponds
to the multiplication of all its incoming messages at the previous
time instant, except the incoming message on the edge that the node
would send back on. If the node of interest is involved in a cycle
of length $L$, it will receive copies of its transmitted monomials,
at different edges, first at $t=L-1$. 

If the node of interest receives $N$ copies of its transmitted monomials,
on different edges, at a certain time instant, $t$, this indicates
that this node is involved in $N$ cycles of length $L=t+1$. These
are referred to as $N_{L}(v_{i})$ or $N_{L}(l_{j})$, depending on
whether the the node of interest is a variable node $v_{i},\forall i=1,\ldots,n$,
or a check node $l_{j},\forall j=1,\ldots,m$, respectively. The process
is repeated for all the nodes within one set of nodes. For example,
looping over the variable nodes for $i=1,\ldots,n$, where $n$ is
the total number of variable nodes, yields the number of cycles, $N_{L}\left(v_{i}\right)$,
of length $L$, that pass through variable node $v_{i}$. After that,
the results are accumulated in a global counter, $C_{L}$, that corresponds
to all the cycles of length $L$, that pass through all the variable
nodes.

\begin{figure}[t]
\begin{centering}
\includegraphics[width=1\columnwidth]{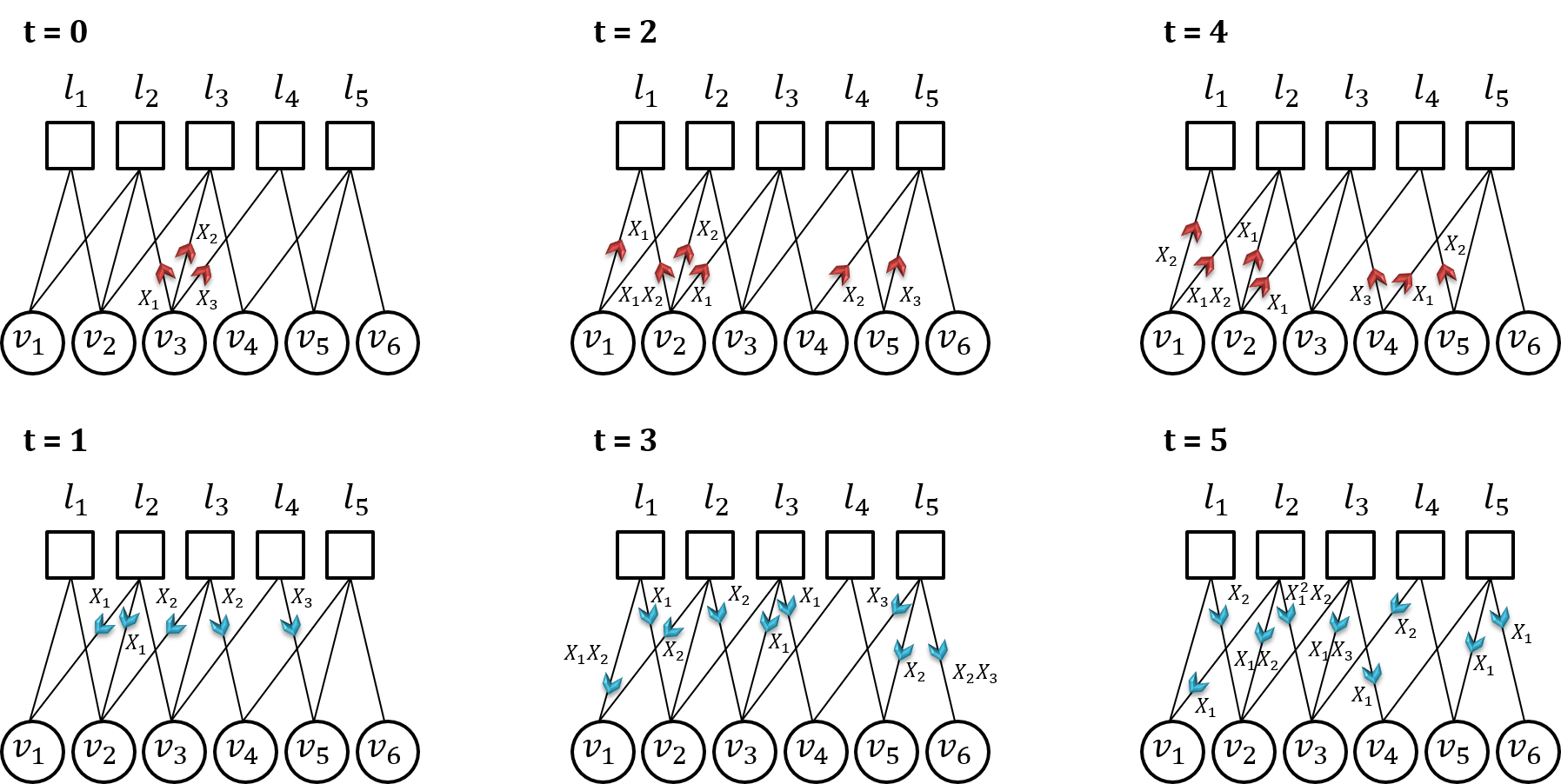}
\par\end{centering}
\caption{Example of Full-Cycle Algorithm on a Tanner graph of $n=6$ variable
nodes and $m=5$ check nodes, counting the cycles that pass through
variable node $v_{3}$}
\label{Full-Cycle Algorithm}

\end{figure}

Figure \ref{Full-Cycle Algorithm} depicts the steps of the Full-Cycle
algorithm, to calculate the cycles that pass through variable node
$v_{3}$. The algorithm is initiated at $t=0$, where $v_{3}$ sends
$X_{1}$, $X_{2}$, and $X_{3}$, along its first, second, and third
edges, respectively. All the other variable nodes would send unities.
At $t=1$, the check nodes gather the information they collect along
their connected edges, in order to calculate the updated information
that they send back. For example, check node $l_{2}$, receives 1,
1, and $X_{1}$, along its first, second, and third edges, respectively.
At $t=1$, $l_{2}$ sends back $X_{1}$, $X_{1}$, and 1, along its
first, second, and third edges, respectively. At $t=2$, and $t=3$,
extrinsic information is passed from variable nodes to check nodes,
and from check nodes to variable nodes, respectively. Because at $t=3$,
$v_{3}$ receives one copy of $X_{2}$ along its first edge, and one
copy of $X_{1}$ along its first edge, one cycle of length $L=t+1=4$
is declared. This means that $N_{4}\left(v_{3}\right)=1$. At $t=4$,
the variable nodes pass their extrinsic information to their connected
check nodes, while $v_{3}$ passes only unities. In return, at $t=5$,
the check nodes update their extrinsic information, and send them
back to their corresponding variable nodes. This time, $v_{3}$ receives
again copies of its monomials at $t=5$, which indicates that there
is one or more cycles of length $L=6$. On its first edge, $v_{3}$
receives $X_{1}^{2}X_{2}$. It discards $X_{1}^{2}$, as it corresponds
to copies of $X_{1}$, which is the initially transmitted monomial
on the first edge of $v_{3}$. This leaves $v_{3}$ with one copy
of $X_{2}$ on its first edge. On the second edge, $v_{3}$ receives
$X_{1}X_{3}$, which corresponds to one copy of $X_{1}$ and one copy
of $X_{3}$, none of which should be discarded. On the third edge,
$v_{3}$ receives a copy of $X_{2}$, which should be kept. In total,
at $t=5$, $v_{3}$ has a copy of $X_{1}$, two copies of $X_{2}$,
and a copy of $X_{3}$. This sums up to four, meaning that the total
number of cycles of length $L=t+1=6$, that pass through $v_{3}$
is $\nicefrac{4}{2}=2$. That is, $N_{6}\left(v_{3}\right)=4$.

Because $v_{3}$ has one cycle of length four, the girth of this Tanner
graph is four. Thus, the algorithm counts all cycles of length up
to $2g-2=6$. For $v_{3}$, the number of 4-cycles, $N_{4}\left(v_{3}\right)$,
and the number of 6-cycles, $N_{6}\left(v_{3}\right)$ should be accumulated
on the global 4-cycle counter, $C_{4}$, and the global 6-cycle counter,
$C_{6}$, respectively. Finally, $v_{3}$ and its three edges should
be deleted from the graph to avoid duplications.

\subsubsection{Half-Cycle Algorithm}

The essence of the Half-Cycle counting algorithm lies in its ability
to get rid of half of the cycle round while adopting the message passing
algorithm to trace the cycles. It debates that because a typical LDPC
cycle on the Tanner graph starts and terminates at the same node,
the cycle could be traced half-way. In essence, the algorithm is able
to count the number of cycles on a given Tanner graph, yielding the
same results as the Full-Cycle algorithm proposed in \cite{karimi2012message},
at half the required computational complexity.

\subsubsection{Cycle Profile Analysis of a Classical LDPC Code}

This section analyzes the cycle profile of the $(4320,2160)$ LDPC
code, proposed in the DVB-NGH standard \cite{gomez2014dvb}. Table
\ref{Cycle Profile NGH} shows the cycle profile of the NGH LDPC code,
obtained from either the Full-Cycle, or the Half-Cycle algorithm,
which yields exactly the same results. Because the Half-Cycle algorithm
requires half the computational complexity of the Full-Cycle algorithm,
in the sequel, only the Half-Cycle algorithm would be utilized. The
girth of the NGH code is found by the algorithm to be six. Consequently,
the algorithm could quantify the number of cycles, whose length is
less than or equal ten. As shown in the table, the analysis indicate
that the NGH code has $C_{6}=31200$, cycles of length six, $C_{8}=1558340$,
cycles of length eight, and $C_{10}=73706359$, cycles of length ten.

\begin{table}[t]
\caption{Cycle Profile of $(4320,2160)$ DVB-NGH LDPC code}

\label{Cycle Profile NGH}%
\begin{tabular}{|c|c|c|c|}
\hline 
 & $C_{6}$ & $C_{8}$ & $C_{10}$\tabularnewline
\hline 
\hline 
$(4320,2160)$ DVB-NGH LDPC & $31200$ & $1558340$ & $73706359$\tabularnewline
\hline 
\end{tabular}

\end{table}

\subsection{SCRAM as a Hybrid Matrix}

There is a one-to-one relationship between the adjacent parity check
matrix of an LDPC code, and its corresponding Tanner graph. Belief
Propagation decoding of LDPC codes incorporates the traverse of soft
information back and forth on the Tanner graph between the variable
nodes and the check nodes. In a similar fashion, for the SCRAM decoder,
the soft information traverses the three-layer Tanner graph back and
forth between variable nodes, and both SA and LDPC check nodes. Inspired
by LDPC, a one-to-one relationship between the SCRAM three-layer Tanner
graph, and an adjacent hybrid matrix that fully describes the joint
decoder, is proposed. 

An $(n,k)$ LDPC code, with $k$ information bits, and $n$ coded
bits, is described by a two-layer Tanner graph that comprises $n$
variable nodes, and $m\geq n-k$ LDPC check nodes. The adjacent parity
check matrix, $\mathbf{H}$, of such code consists of $n$ columns
that correspond to the set of variable nodes, and $m$ rows that correspond
to the LDPC check nodes. A SCRAM system that incorporates $N_{u}$
users, is modeled on a three-layer Tanner graph, with a set of variable
nodes encompassed by two sets of check nodes; SA and LDPC check nodes.
For $n_{u}=1,\ldots,N_{u}$, if user $U_{n_{u}}$ adopts an $(n_{n_{u}},k_{n_{u}})$
LDPC code, then the three-layer Tanner graph comprises a set of $N_{v}=\ensuremath{\sum\limits _{n_{u}=1}^{N_{u}}n_{n_{u}}}$
variable nodes, a set of $N_{l}=\ensuremath{\sum\limits _{n_{u}=1}^{N_{u}}m_{n_{u}}}$
LDPC check nodes, and a set of $N_{s}$ SA check nodes, that correspond
to the $N_{s}$ slots allocated for transmission.

Similar to conventional LDPC codes, the columns of the proposed hybrid
matrix represent the set of variable nodes in the three-layer Tanner
graph. This means that the SCRAM hybrid matrix, $\mathbf{H_{\textrm{SCRAM}}}$,
consists of $N_{v}=\ensuremath{\sum\limits _{n_{u}=1}^{N_{u}}n_{n_{u}}}$
columns, that represent the set of $N_{v}$ variable nodes from the
three-layer Tanner graph. The $i^{\textrm{th}}$ symbol of user $U_{n_{u}}$,
such that $1\leq n_{u}\leq N_{u}$, and $1\leq i\leq n_{n_{u}}$,
corresponds to the $n_{v}^{\textrm{th}}$ column of $\mathbf{H_{\textrm{SCRAM}}}$,
such that $\ensuremath{n_{v}=\ensuremath{\left(\sum\limits _{\acute{n_{u}}=1}^{n_{u}-1}n_{\acute{n_{u}}}\right)+i}}$.

The first $N_{s}$ rows of $\mathbf{H_{\textrm{SCRAM}}}$ correspond
to the $N_{s}$ SA check nodes in the three-layer Tanner graph, or
alternatively the $N_{s}$ frequency subcarriers allocated for transmission.
The population of this part of the matrix depends on the randomly-selected
slots by each of the $N_{u}$ users, to transmit its $n_{n_{u}}$
modulated symbols. For $n_{s}=1,\ldots,N_{s},\;n_{u}=1,\ldots,N_{u}$,
and $i=1,\ldots,n_{n_{u}}$, if user $U_{n_{u}}$ selects SA slot
$s_{n_{s}}$ to transmit its $i^{\textrm{th}}$ modulated symbol,
an entry of 1 is placed at the intersection of row $n_{s}$, and column
$n_{v}$ in $\mathbf{H_{\textrm{SCRAM}}}$, such that $\ensuremath{n_{v}=\ensuremath{\left(\sum\limits _{\acute{n_{u}}=1}^{n_{u}-1}n_{\acute{n_{u}}}\right)+i}}$.

The next set of rows in $\mathbf{H_{\textrm{SCRAM}}}$ corresponds
to the vertical concatenation of the $N_{l}$ LDPC check nodes of
the three-layer Tanner graph. This means that for $n_{u}=1,\ldots,N_{u}$,
the $m_{n_{u}}$ LDPC check nodes of user $U_{n_{u}}$, correspond
to the $m_{n_{u}}$ rows of $\mathbf{H_{\textrm{SCRAM}}}$, that are
located at row indices $N_{s}+\left(\sum\limits _{\acute{n_{u}}=1}^{n_{u}-1}m_{\acute{n_{u}}}\right)+1$
to $N_{s}+\left(\sum\limits _{\acute{n_{u}}=1}^{n_{u}-1}m_{\acute{n_{u}}}\right)+m_{n_{u}}$.
The population of this part of the matrix is deterministic, and depends
fully on the location of ones in the parity check matrix, $\mathbf{H_{\textrm{\ensuremath{n_{u}}}}}$,
of the adopted LDPC code of user $U_{n_{u}}$, for $n_{u}=1,\ldots,N_{u}$.
This means that if user $U_{n_{u}}$ adopts an LDPC code, described
by a parity check matrix, $\mathbf{H_{\textrm{\ensuremath{n_{u}}}}}$,
that has an entry of 1, at the intersection of column $i$, for $i=1,\ldots,n_{n_{u}}$,
and row $j$, for $j=1,\ldots,m_{n_{u}}$, this entry is mapped to
an entry of 1, at the intersection of column $\ensuremath{\ensuremath{\left(\sum\limits _{\acute{n_{u}}=1}^{n_{u}-1}n_{\acute{n_{u}}}\right)+i}}$,
and row $N_{s}+\left(\sum\limits _{\acute{n_{u}}=1}^{n_{u}-1}m_{\acute{n_{u}}}\right)+j$
in $\mathbf{H_{\textrm{SCRAM}}}$.

\begin{figure}[tbh]
\begin{centering}
\includegraphics[width=1\columnwidth]{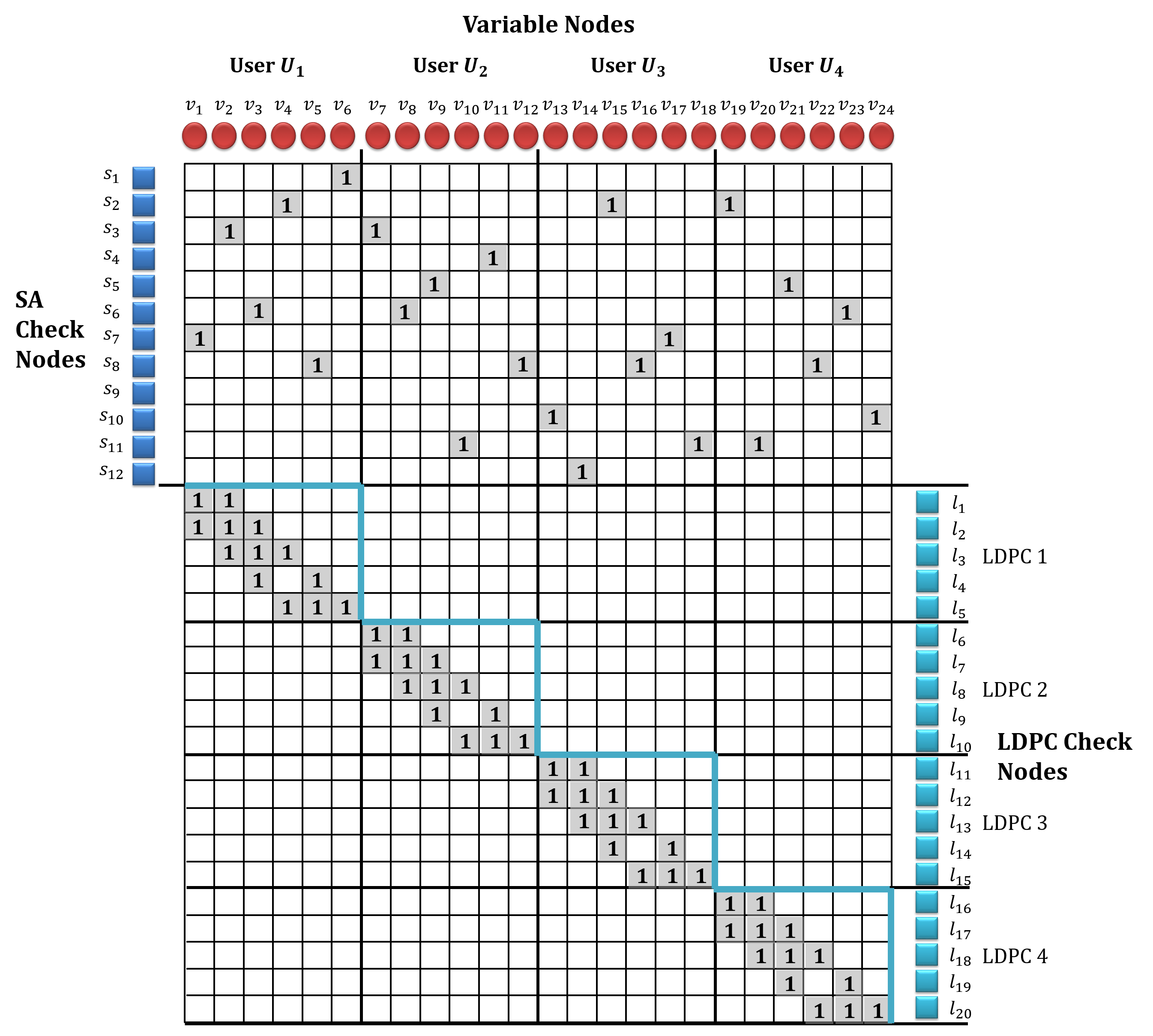}
\par\end{centering}
\caption{Hybrid Matrix of a SCRAM system with $N_{u}=4$ Users, each transmitting
$n_{n_{u}}=6$ LDPC encoded symbols over a system with $N_{s}=12$
SA slots, by means of Random Access}

\label{Hybrid Matrix}
\end{figure}

For illustration, consider again the three-layer Tanner graph example
shown in Figure \ref{Tanner Example}. The hybrid matrix of this model
is depicted in Figure \ref{Hybrid Matrix}. The matrix consists of
$N_{v}=24$ columns representing the variable nodes, and $N_{s}+N_{l}=32$
rows representing both the SA and the LDPC check nodes. The location
of the 1s in the upper SA submatrix (first 12 rows) depends entirely
on the random access selection of the users. For example, the Tanner
graph shows connections between the six modulated symbols of user
$U_{1}$, and SA slots $s_{7}$, $s_{3}$, $s_{6}$, $s_{2}$, $s_{8}$,
and $s_{1}$. This maps to six entries of 1 at the intersection of
the first six columns of $\mathbf{H_{\textrm{SCRAM}}}$, and the corresponding
row indices of the selected SA slots. The blank entries in the matrix
simply denote the absence of connections, and could also be represented
by zeros.

The next 20 rows of $\mathbf{H_{\textrm{SCRAM}}}$ correspond to the
vertical concatenation of the five LDPC check nodes of each of the
four users. Each user possesses connections only to its own set of
LDPC check nodes. This is demonstrated in the block-wise staircase
fashion in the lower LDPC submatrix of $\mathbf{H_{\textrm{SCRAM}}}$.
Within the block zone of a specific user, the entries are regulated
by the location of ones in the parity check matrix of the adopted
LDPC code of that user. For example, the second variable node of user
$U_{3}$ is connected to its first, second, and third LDPC check nodes.
On $\mathbf{H_{\textrm{SCRAM}}}$, this is translated to three entries
of 1 at the intersection of the $14^{\textrm{th}}$ column and the
$23^{\textrm{rd}}$, $24^{\textrm{th}}$, and $25^{\textrm{th}}$
rows.

\subsection{Cycle-Profile of SCRAM\label{subsec:Cycle-Profile-of-SCRAM}}

The essence of mapping the joint SCRAM system to a hybrid matrix representation,
lies in enabling the feeding of the mapped hybrid SCRAM matrix, to
one of the discussed cycle counting algorithms (Full-Cycle \cite{karimi2012message}
or Half-Cycle \cite{li2015improved}), which in return yields both
the girth, and the cycle profile of the mapped SCRAM system. Because
the Half-Cycle algorithm requires only half the computational complexity,
it is chosen as the candidate cycle-counting algorithm of the matrix
representation of the SCRAM Tanner graph.

Table \ref{Cycle Profile SCRAM} depicts the cycle profile of a SCRAM
system that accommodates $N_{u}=4$ users, where each user deploys
the $(4320,2160)$ DVB-NGH \cite{gomez2014dvb} LDPC code. The shared
channel comprises $N_{s}=8640$ frequency subcarriers. The three-layer
Tanner graph is first mapped to hybrid SCRAM matrix, and fed to the
Half-Cycle counting algorithm. The girth of the SCRAM systems was
found by the algorithm to be six. Consequently, the Half-Cycle counting
algorithm is capable of quantifying all short cycles of length up
to ten. However, in order to reduce the computational complexity,
the search was limited to finding only the cycles of length six and
eight, as they are provisioned to have the more dominant impact on
the systems' performance. For convenience, the table also includes
the cycle profile of the classical DVB-NGH LDPC code, that was presented
in Table \ref{Cycle Profile NGH}.

As shown in the table, the SCRAM system possesses $C_{6}=124800$
cycles of length six, and $C_{8}=6234085$ cycles of length eight.
A careful inspection reveals that the total number of $6-$cycles
in the joint SCRAM system, is four times the number of $6-$cycles
in the pure LDPC case. This implies that the only source of $6-$cycles,
on the three-layer Tanner graph of SCRAM, is the local cycles on the
individual LDPC subgraphs of each of the four accommodated users.

\begin{table}[t]
\caption{Cycle Profile of SCRAM, with $N_{u}=4$ users, adopting the $(4320,2160)$
DVB-NGH LDPC code, with Random Access over a channel with $N_{s}=8640$
slots}

\label{Cycle Profile SCRAM}%
\begin{tabular}{|c|c|c|c|}
\hline 
 & $C_{6}$ & $C_{8}$ & $C_{10}$\tabularnewline
\hline 
\hline 
$(4320,2160)$ DVB-NGH LDPC & 31200 & 1558340 & 73706359\tabularnewline
\hline 
SCRAM, with $N_{u}=4$ users & 124800 & 6234085 & \tabularnewline
\hline 
\end{tabular}
\end{table}

\section{Derivation of SCRAM Girth\label{sec:Girth}}

In the following, a step-by-step analysis of the flow of beliefs on
the three-layer Tanner graph of the proposed joint SCRAM decoder is
presented. Without loss of generality, the analysis is illustrated
on the SCRAM example in Figure \ref{Tanner Example}. However, the
findings apply to any arbitrary SCRAM model. Because the purpose of
is finding the girth of the SCRAM Tanner graph, during the analysis,
any potential cycle with provisioned cycle length that exceeds the
girth (so far found cycle with minimum length) will be discarded.

Two sub-definitions of cycles of the three-layer Tanner graph are
declared; a local cycle and a global cycle. A local cycle on the SCRAM
Tanner graph denotes a cycle that initiates and terminates at a given
variable node, that belongs to a certain user, by means of traversing
a path between variable nodes and LDPC check nodes that represent
the LDPC code of the denoted user. A global cycle on the other hand,
is defined as a cycle that initiates and terminates at a variable
node, that belongs to a certain user, by means of traversing a path
between variable nodes and LDPC check nodes that belong to different
users, via trespassing their commonly connected SA check nodes.

Both the local cycles and global cycles are considered as cycles of
the three-layer Tanner graph. Consequently, the overall girth of the
SCRAM graph, is counted as the minimum cycle length of all the local
and global cycles. The Half-Cycle algorithm could easily identify
the girth of the local cycles. This can be achieved by passing the
parity check matrices of the underlying LDPC codes of the $N_{u}$
users, one at a time. For $n_{u}=1,\ldots,N_{u}$, let $g_{n_{u}}$
be the calculated girth of the underlying LDPC code of user $U_{n_{u}}$,
at the output of the Half-Cycle algorithm. The least local cycle length,
$g_{local}$, is thus the minimum of all the $N_{u}$ girths of the
different LDPC codes. That is, $g_{local}=\ensuremath{\min\limits _{n_{u}}\left(g_{n_{u}}\right)}$. 

More generally, passing the SCRAM hybrid matrix, $\mathbf{H_{\textrm{SCRAM}}}$,
to the Half-Cycle algorithm could yield the overall girth, $g_{SCRAM}$,
of the three-layer Tanner graph. However, the complexity of the algorithm
grows extensively with the number of users, and the codeword length.
The incentive of the analysis proposed herein, is to derive a lower-bound
on the length, $g_{global}$, of the shortest global cycle in the
graph. The overall girth, $g_{SCRAM}$, of the SCRAM graph would then
be the minimum of the local girth, $g_{local}$, and the analyzed
global girth, $g_{global}$.

Consider again the SCRAM example in Figure \ref{Tanner Example},
with $N_{u}=4$ users, $n_{n_{u}}=6$ symbols per user, $m_{n_{u}}=5$
LDPC check nodes per user, and $N_{s}=12$ SA slots. As aforementioned,
the four users adopt identical LDPC codes. The girth of each of the
adopted LDPC codes was found to be four. That is, $g_{n_{u}}=4\;\forall n_{u}=1,\ldots,4$.
Thus, the minimum local girth is given by $g_{local}=4$. In the following,
the example is analyzed to track the shortest length of global cycles.
For illustration, variable node $v_{3}$ is chosen as the initiating
node of the belief. Hence, a full global cycle is declared when the
belief traverses the three-layer Tanner graph all the way back to
$v_{3}$. Equivalently, any of the $N_{v}=24$ variable nodes is a
suitable candidate for the initiation of the message.

\begin{figure}[t]
\begin{centering}
\includegraphics[width=1\columnwidth]{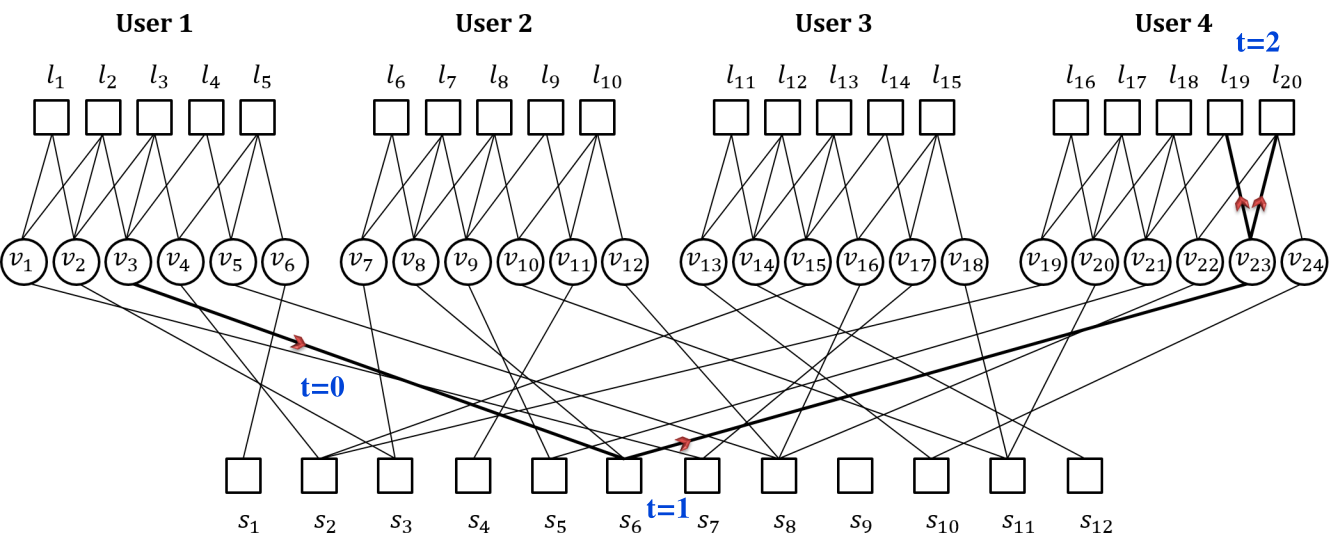}
\par\end{centering}
\caption{Track of flow of beliefs at $t=0,1,2$, on a SCRAM graph with $N_{u}=4$
users, $n_{n_{u}}=6$ symbols, and $N_{s}=12$ slots}
\label{girth_t012}
\end{figure}

Figure \ref{girth_t012} depicts in bold the initialization step at
time instant $t=0$, such that only variable node $v_{3}$ is active,
and sends a monomial (e.g. $X_{1}$) concurrently in both directions
to its adjacent SA check node, $s_{6}$, and LDPC check nodes, $l_{2}$,
$l_{3}$, and $l_{4}$. The monomial transmitted to the LDPC check
nodes has two provisioned paths in the subsequent time instants. The
first path is a local cycle between the variable nodes and LDPC check
nodes that belong to the subgraph of user $U_{1}$. The girth of such
a local cycle can not go below the pre-calculated local girth, $g_{1}$,
of the underlying LDPC code of user $U_{1}$. The second provisioned
path of the monomials received at the LDPC check nodes, is that in
the next time slot they flow form the LDPC check nodes to other variable
nodes that belong to user $U_{1}$. For example, LDPC check node,
$l_{2}$, would forward the monomial to variable nodes $v_{1}$, and
$v_{2}$, which in return will forward the monomial to their respective
SA check nodes. The global cycle analysis in this case would translate
to finding the length of the cycles initiated from either $v_{1}$
or $v_{2}$, in addition to the two extra time slots (or transitions),
required for the monomial to hop from $v_{3}$ to $l_{2}$ till it
lands on $v_{1}$ and $v_{2}$. Because the main interest is to identify
the shortest cycle length, this provisioned path is discarded as it
yields a cycle of longer length. The only relevant transition at this
time instant, is thus the transition from variable node $v_{3}$ to
SA slot $s_{6}$. Hence, for the subsequent analysis, the cycle will
be traced further from $s_{6}$.

At $t=1$, SA check node $s_{6}$ forwards the monomial it received
from variable node $v_{3}$, at the previous time instant, to its
adjacent variable nodes, $v_{8}$, and $v_{23}$. The concept of extrinsic
information prohibits $s_{6}$ from transmitting the monomial back
to $v_{3}$. Because the purpose is to analyze and find the shortest
path, the idea could be fulfilled by means of tracking either $v_{8}$
or $v_{23}$. This statement would have been misleading, if the target
was to calculate the number of cycles of length that is equal to the
girth. However, for identifying the girth, considering one variable
node only would be sufficient. Thus, in order to avoid confusion,
for the rest of the analysis, the cycle is to be traced from $v_{23}$.

Figure \ref{girth_t012} also shows the transition of the monomial
from variable node $v_{23}$ to its adjacent LDPC check nodes $l_{19}$
and $l_{20}$ at $t=2$. Again, due to extrinsic message passing,
the monomial can not be sent back from $v_{23}$ to SA check node
$s_{6}$. Proceeding from this point, two different paths with different
properties could be taken depends on whether the cycle is traced from
$l_{19}$ or from $l_{20}$. These two possibilities should be handled
separately to avoid confusion. First, the path from $l_{20}$ would
be tracked for the subsequent time slots. The analysis would track
the monomial after it is forwarded from $l_{20}$, all the way till
it reaches $v_{3}$. After that, the analysis would go back to this
point in time, and track the path taken from $l_{19}$. These two
cases of starting from $l_{20}$, and starting from $l_{19}$, would
be denoted as case 1 and case 2, respectively.

\begin{figure}[t]
\begin{centering}
\includegraphics[width=1\columnwidth]{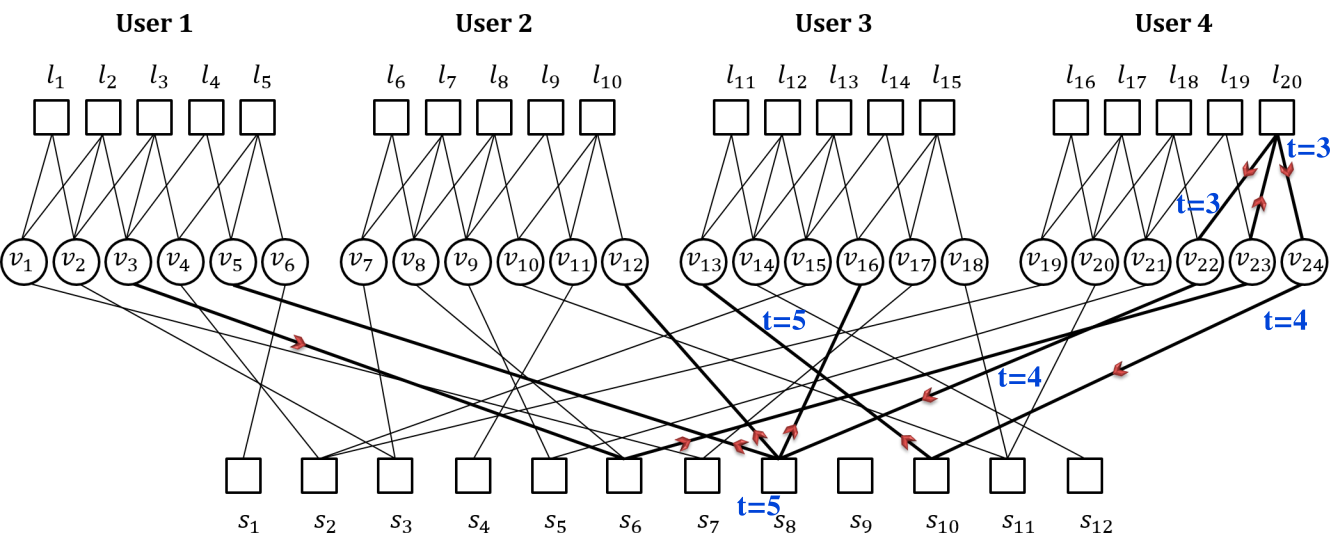}
\par\end{centering}
\caption{Track of flow of beliefs at $t=3,4,5$, Case 1, on a SCRAM graph with
$N_{u}=4$ users, $n_{n_{u}}=6$ symbols, and $N_{s}=12$ slots}
\label{girth_t345_C1}
\end{figure}

At $t=3$, as shown in Figure \ref{girth_t345_C1}, for case 1, LDPC
check node, $l_{20}$, forwards the monomial to its adjacent variable
nodes $v_{22}$ and $v_{24}$. These also yield different paths with
different features that would be handled jointly. Consequently, for
the next time slots, the analysis would continue from both $v_{22}$
and $v_{24}$ simultaneously.

Figure \ref{girth_t345_C1} also depicts the forwarding of the monomials,
at $t=4$, from variable nodes $v_{22}$ and $v_{24}$ to their adjacent
SA check nodes $s_{8}$ and $s_{10}$, respectively. Additionally,
variable node $v_{22}$ should also forward the monomial to its adjacent
LDPC check node $l_{18}$. However, the path from $l_{18}$ would
either form a local LDPC cycle, or a global cycle of two extra transitions.
Consequently, only the transitions to SA check nodes $s_{8}$ and
$s_{10}$ are relevant, and would be kept for further analysis.

As shown in Figure \ref{girth_t345_C1}, at $t=5$, $s_{8}$ forwards
the monomial it received from $v_{22}$ to its adjacent variable nodes
$v_{5}$, $v_{12}$, and $v_{16}$. Meanwhile, $s_{10}$ forwards
the monomial it received from $v_{24}$ to its adjacent variable node,
$v_{13}$. Considering the transitions of $s_{8}$, variable node
$v_{5}$ belongs to user $U_{1}$. This is the same user that accommodates
$v_{3}$, the initiating variable node of the monomial at $t=0$.
Since $v_{5}$ and $v_{3}$ belong to the same user, if they share
a common LDPC check node, the monomial would only need two more transitions
to flow from $v_{5}$ to $v_{3}$, via their common LDPC check node.
Thus, the path from $v_{5}$ is considered as a candidate path of
the shortest cycle analysis. Conversely, $v_{12}$ and $v_{16}$ belong
to users $U_{2}$ and $U_{3}$, respectively. This means that they
do not share any common LDPC check nodes with $v_{3}$. Hence, the
monomial would need more than two transitions to flow from $v_{12}$
and $v_{16}$ to $v_{3}$. It is also guaranteed that $v_{3}$ does
not share the same SA check node with $v_{12}$ and $v_{16}$, because
otherwise, the monomial initiated by $v_{3}$ at $t=0$, would have
already landed on $v_{12}$ and $v_{16}$ at $t=1$, which is not
the case. Hence, the paths from $s_{8}$ to both $v_{12}$ and $v_{16}$
would require more transitions to reach $v_{3}$, and are thus excluded
from the candidate set. The same argument could be made for the path
from $s_{10}$ to $v_{13}$, which belongs to the variable nodes of
user $U_{3}$. In conclusion, the only relevant path at this time
instant is the transition from $s_{8}$ to $v_{5}$.

\begin{figure}[t]
\begin{centering}
\includegraphics[width=1\columnwidth]{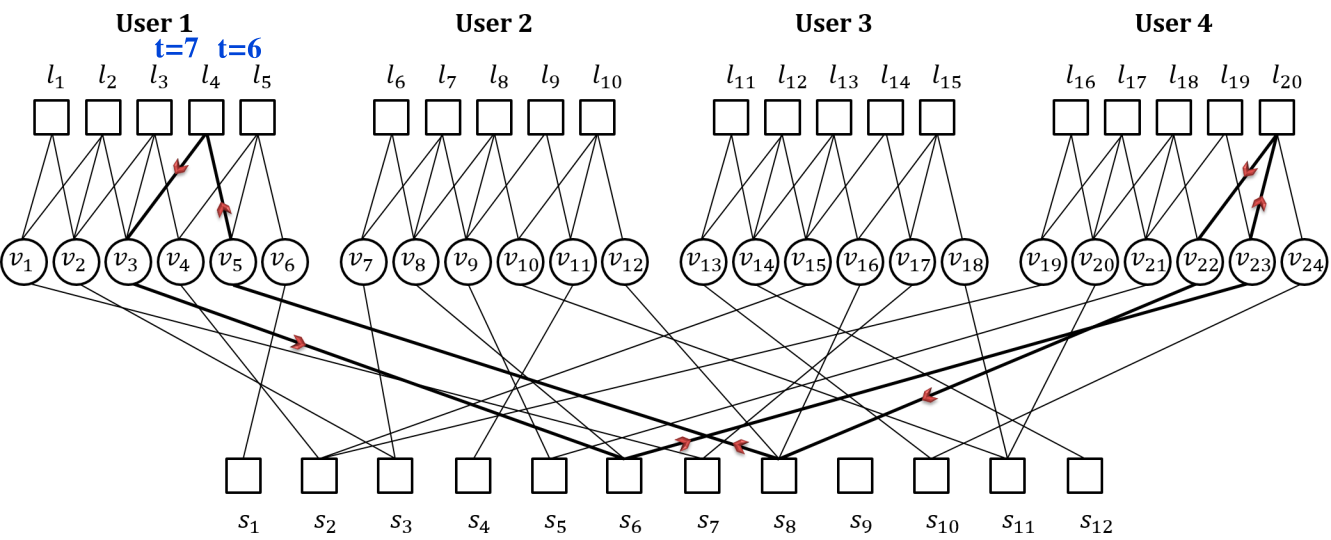}
\par\end{centering}
\caption{Track of flow of beliefs at $t=6,7$, Case 1, on a SCRAM graph with
$N_{u}=4$ users, $n_{n_{u}}=6$ symbols, and $N_{s}=12$ slots}
\label{girth_t67_C1}
\end{figure}

The transition from $v_{5}$ to LDPC check nodes $l_{4}$ and $l_{5}$,
at $t=6$, is shown in Figure \ref{girth_t67_C1}. As shown in the
figure, $l_{4}$ is directly connected to $v_{3}$. This means that
at the next time instant, $l_{4}$ would forward the monomial to $v_{3}$.
On the other hand, $l_{5}$ does not exhibit a direction connection
to $v_{3}$. This means that $l_{5}$ would require at least two more
time instants in comparison to $l_{4}$, in order to send the monomial
back to $v_{3}$. As a result, the path from $l_{5}$ is excluded
from the candidate set of the shortest cycle.

Finally, at $t=7$, as shown in Figure \ref{girth_t67_C1}, $l_{4}$
forwards the monomial to $v_{3}$. This completes one global cycle
of eight transitions, that starts and terminates at variable node
$v_{3}$. In other words, the monomial transmitted by variable node
$v_{3}$, traverses the three-layer Tanner graph between SA check
nodes, and variable and check nodes of the remaining users, and returns
back to $v_{3}$ after eight transitions, resulting in a global cycle
of length eight. 

\begin{figure}[b]
\begin{centering}
\includegraphics[width=1\columnwidth]{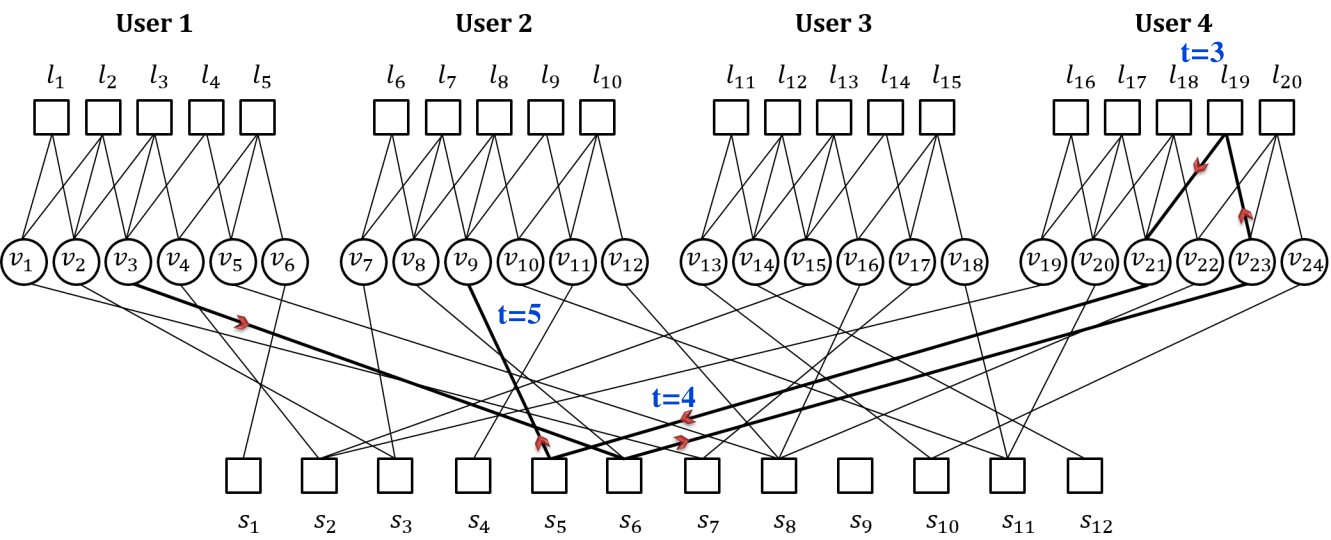}
\par\end{centering}
\caption{Track of flow of beliefs at $t=3,4,5$, Case 2, on a SCRAM graph with
$N_{u}=4$ users, $n_{n_{u}}=6$ symbols, and $N_{s}=12$ slots}
\label{girth_t345_C2}
\end{figure}

Now consider the second case left out at $t=2$, where the monomial
arrives at LDPC check node, $l_{19}$. As shown in Figure \ref{girth_t345_C2},
at $t=3$, $l_{19}$ would forward the monomial to $v_{21}$. In return,
at $t=4$, $v_{21}$ would forward it to SA check node $s_{5}$, and
to its adjacent LDPC check nodes, $l_{17}$ and $l_{18}$. The forwarded
monomial to the LDPC check nodes would either make a local cycle,
or proceed further to a global cycle of two extra transitions, in
comparison to the path to $s_{5}$. Consequently, the paths to the
LDPC check nodes are discarded, leaving only the transition to $s_{5}$.
At $t=5$, $s_{5}$ forwards the monomial to variable node, $v_{9}$,
which belongs to user $U_{2}$. Similar to the previous discussion
in the first case, because $v_{9}$ and $v_{3}$ belong to different
users, $v_{9}$ would require more than two more transitions to send
the monomial back to $v_{3}$. This means that the least cycle length
due to this path would exceed eight, which is the length of the cycle
traced from $l_{20}$ in the first case. Therefore, case 1 is declared
as the shortest global cycle, with a cycle length of eight. 

\begin{figure}[t]
\begin{centering}
\includegraphics[width=1\columnwidth]{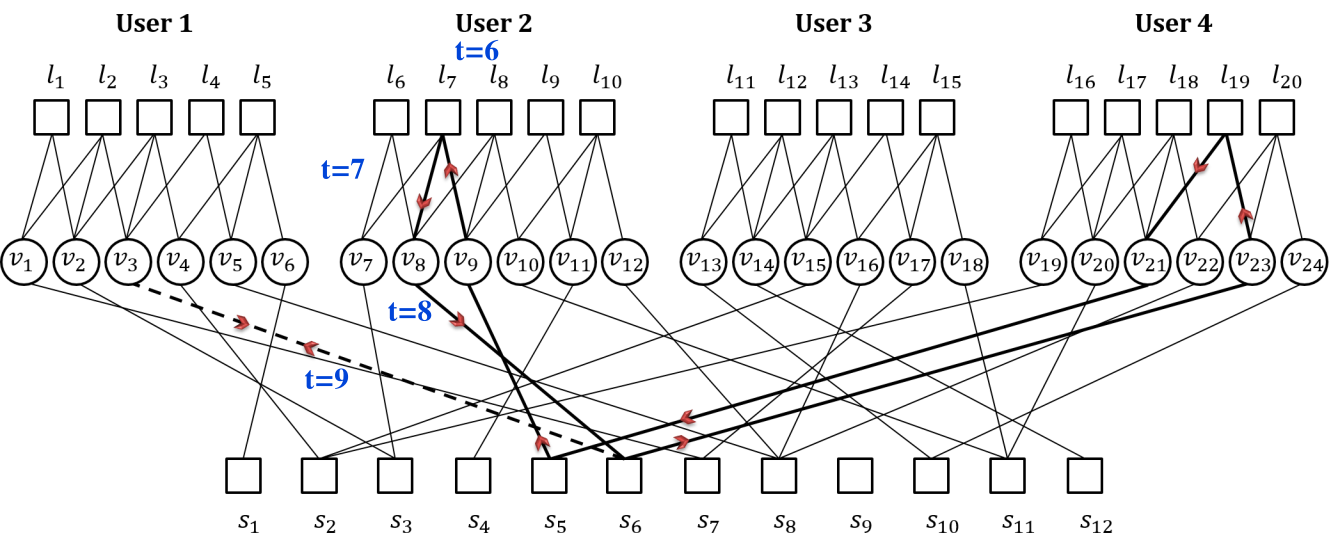}
\par\end{centering}
\caption{Example of a closed walk at $t=9$, Case 2, on a SCRAM graph with
$N_{u}=4$ users, $n_{n_{u}}=6$ symbols, and $N_{s}=12$ slots}
\label{closedWalk}
\end{figure}

Tracing the second case further from $v_{9}$ to $l_{7}$ to $v_{8}$
to $s_{6}$ to $v_{3}$ shows an example of a closed walk as depicted
in Figure \ref{closedWalk}. Although the monomial returns back to
$v_{3}$ after ten transitions, the walk is not considered a cycle
because the starting and ending edges are the same.

\begin{figure}[b]
\begin{centering}
\includegraphics[width=1\columnwidth]{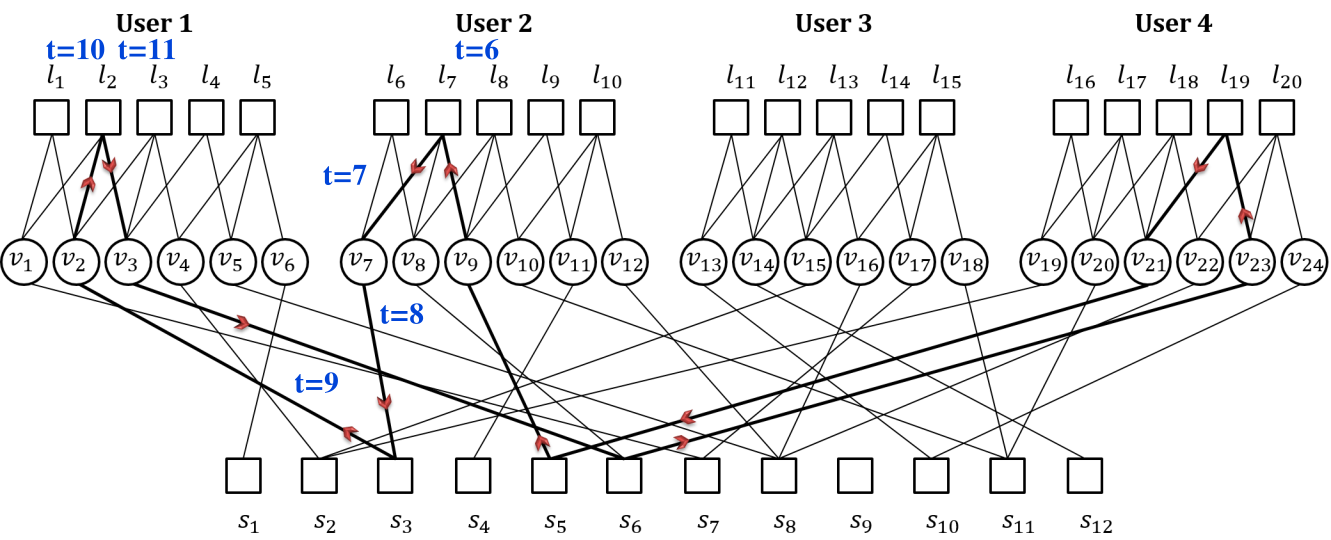}
\par\end{centering}
\caption{Track of flow of beliefs at $t=6,7,8,9,10,11$, Case 2, on a SCRAM
graph with $N_{u}=4$ users, $n_{n_{u}}=6$ symbols, and $N_{s}=12$
slots}
\label{girth_t11_C2}
\end{figure}

The example of the closed walk leads to an important observation,
that is depicted in Figure \ref{girth_t11_C2}. The figure shows the
further tracking of the monomial after it arrives at $v_{9}$ at $t=5$,
till it reaches $v_{3}$, after a global cycle of length 12. This
means that this part of the analysis considers the case where the
monomial at $v_{9}$ would follow a different path, than the path
that lead to the closed walk. For this path, the monomial is forwarded
from $v_{9}$ to $l_{7}$ at $t=6$, and from $l_{7}$ to $v_{7}$
at $t=7$. After that, the monomial is forwarded from $v_{7}$ to
$s_{3}$, and from $s_{3}$ to $v_{2}$, at $t=8$ and $t=9$, respectively.
The major observation here is that due to the fact that after six
transitions from the initial phase at $t=0$, the monomial landed
on a variable node ($v_{9}$), that does not belong to the variable
node set of the initiating variable node ($v_{3}$), the monomial
required at least four more transitions ($t=6$\textrightarrow $t=9$),
in order to be able to leave the variable node set of the variable
node ($v_{9}$) that it landed on, and land on a variable node ($v_{2}$),
that belongs to the variable node set of the initiating variable node
($v_{3}$). Now because $v_{2}$ and $v_{3}$ have a common LDPC check
node ($l_{2}$), after two more time instants, that is at $t=11$,
the monomial would arrive at $v_{3}$ via passing by $l_{2}$. Consequently,
the total cycle length of this global cycle is $6+4+2=12$. Any other
attempt to reach $v_{3}$ from $v_{9}$, would either go through a
closed walk, or a global cycle of length at least 12.

To wrap up the analysis, the first six time slots of a global cycle
involve three crucial pair of transitions. The first pair represents
the transition from the initiating variable node that belongs to a
certain user, to a second variable node that belongs to a different
user, via passing by a common SA check node. The second pair involves
a transition from the second variable node, to another variable node
that belongs to the same user, via passing by a shared LDPC check
node. For the third pair of transitions, the monomial is forwarded
from the receiving variable node from the previous pair, to another
variable node in a different user set, via passing by a common SA
check node. These three pairs of transitions are crucial, and require
six time slots. Two possible scenarios could arise after the landing
on the last variable node from the last pair. 

The first scenario corresponds to the case where the landed on variable
node belongs to the same user of the initiating variable node at $t=0$.
In this case, if the two variable nodes share a common LDPC check
node, then only one pair of transitions is required for the information
to go back to its initiating variable node. Consequently, the total
cycle length of the global cycle is eight. If on the other hand, the
two variable nodes do not have a common LDPC check node, then at least
two more pairs of transitions are required to complete the cycle.
In this case, the cycle length is at least 10. 

The second scenario corresponds to the case where the landed on variable
node does not belong to the same user as the initiating variable node.
In this case, as was discussed in the last example, at least two more
pairs of transitions are required for the monomial to leave the current
user set, and land on a variable node within the user set of the initiating
variable node. After these two more pairs, the situation translates
to the first scenario, which means another one or more transitions
is required to finish the cycle. As a result, this situation requires
at least three more pairs of transitions, in addition to the crucial
three transitions. Consequently, the least cycle length from the second
scenario is 12.

From the above discussion, it can be deduced that the shortest length
of a global cycle on the SCRAM Tanner graph is at least eight. This
indicates that the global girth is given by $g_{global}\geq8$. If
any of the adopted LDPC codes, has a local girth, $g_{local}$, that
is less than eight, the overall SCRAM girth would match the smaller
girth, whether it is the local girth of the individual LDPC codes,
or the global girth of at least eight. Consequently, it can be deduced
that the overall SCRAM girth is given by $g_{SCRAM}\geq\textrm{min}\left(8,g_{local}\right)$. 

\section{A Proposed Graphical Correlation-Based Method for Counting the Number
of Global 8-Cycles on the SCRAM Three-Layer Tanner Graph}

As mentioned previously, the mapping of the SCRAM three-layer Tanner
graph to a hybrid matrix enables the utilization of the Full-Cycle
or the Half-Cycle algorithms, initially proposed for classical LDPC
codes, in computing the cycle profile of SCRAM. However, the complexity
of these algorithms grows exponentially in the number of users and
the number of LDPC symbols per user. In the previous section, it was
shown that the least global cycle length on a SCRAM three-layer Tanner
graph was found to be eight. In this section, a graphical method that
counts the number of global 8-cycles on the SCRAM three-layer Tanner
graph is proposed. Instead of computing the full-cycle profile, the
proposed algorithm focuses on the global 8-cycles, as they are provisioned
to have a more detrimental effect on the convergence of the joint
SCRAM decoding.

As discussed in the previous section, a global cycle of length eight,
on the SCRAM three-layer Tanner graph, involves four transition pairs.
The first pair represents the hopping from the initiating variable
node of one user, to a variable node that belongs to a different user,
via a common SA check node. The second pair corresponds to a hop from
the receiving variable node, to another variable node within its user
set, via a common LDPC check node. The third pair involves the return
trip from the last variable node, back to another variable node within
the set of the initiating user, via a common SA check node. Finally,
the fourth pair corresponds to the hopping from the receiving variable
node, to the initiating variable node, via a common LDPC check node.
For simplicity, in the sequel, two variable nodes, that belong to
the same user, are referred to as connected variable nodes, if and
only if there is a common LDPC check node, that is simultaneously
connected to each one of them, via a single edge. With this definition,
it can be summed up that a global cycle of length eight involves two
connected variable nodes on one user set, two connected variable nodes
on a different user set, and two common SA check nodes, shared by
the four variable nodes.

In order to count the total number of global 8-cycles, the algorithm
loops over all the variable nodes, and calculates the number of global
8-cycles that pass through each one of them. For every iteration,
the user that accommodates the variable node of interest is referred
to as the primary user, while all the remaining users are referred
to as the secondary users. In addition, the variable node of interest
is denoted by $v_{1}^{\left(p\right)}$, which denotes the first primary
variable node. Because a global 8-cycle involves two connected variable
nodes on one user set, the connected variable node to $v_{1}^{\left(p\right)}$
is denoted by $v_{2}^{\left(p\right)}$, and their connecting LDPC
check node is referred to as $l^{\left(p\right)}$. In a similar fashion,
for the second connected pair of variable nodes of the global 8-cycle,
the first secondary variable node, the second secondary variable node,
and their connecting LDPC check node, on the secondary user set, are
referred to as, $v_{1}^{\left(s\right)}$, $v_{2}^{\left(s\right)}$,
and $l^{\left(s\right)}$, respectively. Finally the shared SA check
node that connects the first primary- and the first secondary variable
nodes is denoted by $s_{1}^{\left(p,s\right)}$. Similarly, the shared
SA check node that connects the second primary- and the second secondary
variable nodes is denoted by $s_{2}^{\left(p,s\right)}$.

For a SCRAM three-layer Tanner graph with $N_{v}$ variable nodes,
the proposed algorithm calculates the number of global 8-cycles, $C_{8}^{\left(v_{n_{v}}\right)}$
that pass through variable node $v_{n_{v}},\;\forall n_{v}=1,\cdots,N_{v}$,
and accumulates it to the global 8-cycle counter, $C_{8}^{\left(global\right)}.$
For every iteration, the first primary variable node, $v_{1}^{\left(p\right)}$
is assigned to $v_{n_{v}}$. Let $A_{v_{1}^{\left(p\right)}}^{\left(L\right)}$
, be the set of LDPC check nodes, connected to the first primary variable
node. While the outer loop of the algorithm loops over the variable
nodes, the first inner loop should loop over their adjacent LDPC check
nodes. For every iteration on the set of adjacent LDPC check nodes,
the check node of interest is referred to as $l^{\left(p\right)}$,
which denotes the primary LDPC check node. At this stage, the algorithm
starts a new nested loop, that iterates over the edges of the primary
LDPC check node. Let $A_{l^{(p)}}$, denote the set of variable nodes
connected to the primary LDPC check node, $l^{\left(p\right)}$. Moreover,
let $A_{l^{(p)}/v_{1}^{(p)}}$, denote the set of adjacent variable
nodes of the primary LDPC check node, except the first primary variable
node. For every iteration over $A_{l^{(p)}/v_{1}^{(p)}}$, the selected
variable node is referred to as $v_{2}^{\left(p\right)},$which denotes
the second primary variable node.

Now that $v_{1}^{(p)}$ and $v_{2}^{(p)}$ have been identified, the
next step is to identify their adjacent SA check nodes. Let $A_{v_{1}^{(p)}}^{(S)}$,
and $A_{v_{2}^{(p)}}^{(S)}$, be the set of SA check nodes, connected
to $v_{1}^{(p)}$ and $v_{2}^{(p)}$, respectively. Without loss of
generality, due to the fact that every modulated symbol is transmitted
only once, each of the two sets includes a single SA check node. This
means that the first constituent SA check node, $s_{1}^{(p,s)}$,
and the second constituent SA check node, $s_{2}^{(p,s)}$, are the
only elements in $A_{v_{1}^{(p)}}^{(S)}$ and $A_{v_{2}^{(p)}}^{(S)}$,
respectively.

To complete the constituents of the global $8-$cycle, the two secondary
variable nodes are to be identified. These variable nodes should be
colliding on the pair of constituent SA check nodes, $s_{1}^{(p,s)}$,
and $s_{2}^{(p,s)}$. Moreover, they both should belong to the same
secondary user. Furthermore, they should be connected via a common
LDPC check node. As a first step, all the colliding variable nodes
on $s_{1}^{(p,s)}$, and $s_{2}^{(p,s)}$ have to be identified. Let
$A_{s_{1}^{(p,s)}}$, and $A_{s_{2}^{(p,s)}}$, denote the set of
variable nodes, colliding at $s_{1}^{(p,s)}$, and $s_{2}^{(p,s)}$,
respectively. Because the primary variable nodes have to be excluded
from the potential candidates of secondary variable nodes, let $A_{s_{1}^{(p,s)}/v_{1}^{(p)}}$,
and $A_{s_{2}^{(p,s)}/v_{2}^{(p)}}$, denote the set of variable nodes
that collide at $s_{1}^{(p,s)}$, excluding $v_{1}^{(p)}$, and the
set of variable nodes that collide at $s_{2}^{(p,s)}$, excluding
$v_{2}^{(p)}$, respectively. At this stage, the algorithm loops over
the variable nodes in $A_{s_{1}^{(p,s)}/v_{1}^{(p)}}$, and $A_{s_{2}^{(p,s)}/v_{2}^{(p)}}$,
and for every iteration denotes the selected variable node from $A_{s_{1}^{(p,s)}/v_{1}^{(p)}}$
and $A_{s_{2}^{(p,s)}/v_{2}^{(p)}}$ as the first secondary variable
node, $v_{1}^{(s)}$, and the second secondary variable node, $v_{2}^{(s)}$,
respectively. 

If the two secondary variable nodes do not belong to the same user,
their combination should be excluded. On the other hand, if they belong
to the same user set, the algorithm proceeds to check if they are
connected, and if so, how many LDPC check nodes they have in common.
This can be done via analyzing the parity check matrix of the underlying
LDPC code. When two variable nodes, share a common LDPC check node,
the parity check matrix possesses a value of one, at the row intersection
of the respective LDPC check node, and the two column indices of the
two variable nodes. Thus, the element-wise multiplication, of the
two corresponding columns of the variable nodes of interest, would
yield ones, at the row indices of the LDPC check nodes that the two
variable nodes are simultaneously connected to. Moreover, the summation
of the elements within the resulting column after the multiplication,
yields the total number of LDPC check nodes, that the two variable
nodes share in common. Each common LDPC check node, represents a potential
transition pair between the two variable nodes, to close the global
$8-$cycle. In other words, for every common LDPC check node between
the secondary variable nodes, the cycle counter is incremented by
one. If on the other hand, the two secondary variable nodes, do not
share any common LDPC check nodes, then this pair of secondary variable
nodes is excluded from the potential $8-$cycle constituent candidates.

It is worth pointing out that the indices of the variable nodes, and
LDPC check nodes within the joint three-layer Tanner graph, are different
from their corresponding indices within the underlying parity check
matrix of the adopted LDPC code. For that reason, a simple mapping
between the indices on the joint Tanner graph, and the respective
indices within the underlying LDPC code, has to be made. Assuming,
that all the $N_{u}$ users, adopt identical LDPC codes, with $n_{n_{u}}$
variable nodes, and $m_{n_{u}}$ LDPC check nodes each, then for $n_{v}=1,\ldots,N_{v}$,
such that $N_{v}$ is the total number of variable nodes in the joint
Tanner graph, the corresponding user index, $n_{u}$, to which variable
node, $v_{n_{v}}$, belongs, is given by $n_{u}=\lceil\frac{n_{v}}{n_{n_{u}}}\rceil$.
Moreover, the respective index, $i$, of the variable node within
the underlying LDPC code can be computed as, $i=\mod\left(n_{v}-1,\;n_{n_{u}}\right)+1$.
This means that variable node, $v_{n_{v}}$, within the joint Tanner
graph, corresponds to variable node, $v_{i}$, within the underlying
LDPC code, of user, $U_{n_{u}}$. Similarly, for $n_{l}=1,\ldots,N_{l}$,
such that $N_{l}$ represents the total number of LDPC check nodes
in the joint Tanner graph, LDPC check node, $l_{n_{l}}$, within the
joint Tanner graph, corresponds to LDPC check node, $l_{j}$, within
the underlying LDPC code of user, $U_{n_{u}}$, such that, $n_{u}=\lceil\frac{n_{l}}{m_{n_{u}}}\rceil$,
and $j=\mod\left(n_{l}-1,\;m_{n_{u}}\right)+1$. For the case of non-identical
LDPC codes, the mapping is straightforward, taking into consideration
the respective number of variable nodes, $n_{n_{u}}$, and LDPC check
nodes, $m_{n_{u}}$, within the parity check matrix, $\mathbf{H_{\textrm{LDPC}}^{\left(\textrm{\ensuremath{n_{u}}}\right)}}$,
of user, $U_{n_{u}}$. 

For every common LDPC check node between, $v_{1}^{(s)}$ and $v_{2}^{(s)}$,
the $8-$cycle counter, $C_{8}^{(v_{1}^{\left(p\right)})}$, of the
global $8-$cycles that pass through the first primary variable node,
$v_{1}^{\left(p\right)}$, should be incremented by one. At this point,
the algorithm reverts to the second inner loop, checking if there
are further variable nodes, connected to the primary LDPC check node,
in the set, $A_{l^{(p)}/v_{1}^{(p)}}$, of adjacent variable nodes,
of the current primary LDPC check node, $l^{\left(p\right)}$, excluding
the current first primary variable node, $v_{1}^{\left(p\right)}$.
After terminating the second inner loop, the algorithm proceeds to
check if there are further iterations required in the first inner
loop. This means that the algorithm has to check whether or not there
are further LDPC check nodes, connected to the first primary variable
node. Before proceeding to the outer loop, to investigate a new primary
variable node, the total $8-$cycle counter, $C_{8}^{(global)}$,
should be incremented by the $8-$cycle counter, $C_{8}^{(v_{1}^{\left(p\right)})}$,
of the first primary variable node, $v_{1}^{\left(p\right)}$. A crucial
consideration for the next iterations in the outer loop, is to exclude
any potential $8-$cycle, that involves the previously investigated
primary variable nodes, in order to avoid duplications.

\begin{figure}[tbh]
\begin{centering}
\includegraphics[width=1\columnwidth]{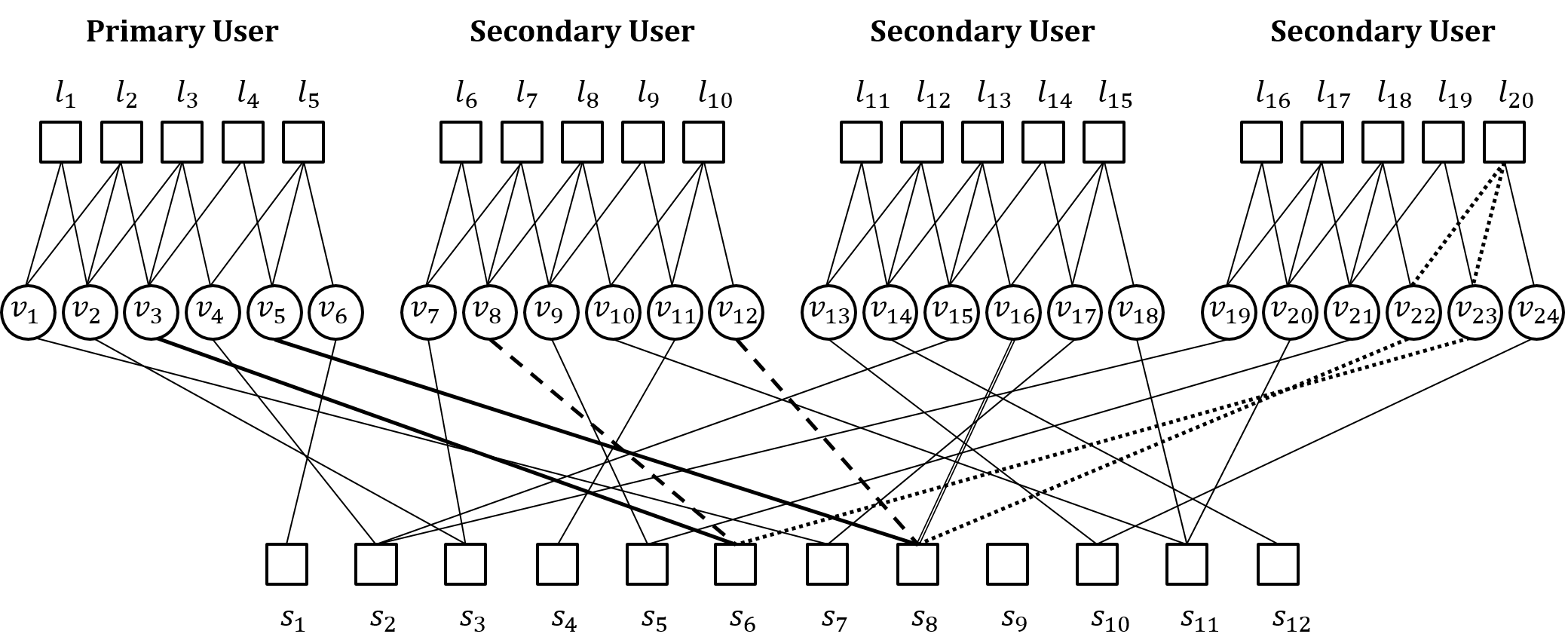}
\par\end{centering}
\caption{Example of algorithmic tracking of $8-$cycles between four users
on a SCRAM Tanner graph with twelve SA slots, Case 2}
\label{algorithmic calculation-arbitrary access-4 users- with cycles}
\end{figure}

For illustration, consider the example shown in Figure \ref{algorithmic calculation-arbitrary access-4 users- with cycles},
with four users, adopting the same LDPC codes as the previous examples.
The system incorporates 12 channel access slots, and the users access
the channel by means of random access. Assume that the objective is
to track the potential global $8-$cycles, that pass through variable
nodes $v_{3}$ and $v_{5}$, that belong to user $U_{1}$. As a result,
user $U_{1}$, is denoted by the primary user, while users, $U_{2}$,
$U_{3}$, and $U_{4}$, are referred to as the secondary users. Moreover,
the first and second primary variable nodes, $v_{1}^{(p)}$ and $v_{2}^{(p)}$,
are respectively assigned to $v_{3}$ and $v_{5}$. Consequently,
the first and second constituent SA check nodes, $s_{1}^{(p,s)}$
and $s_{2}^{(p,s)}$, should be assigned to $s_{6}$ and $s_{8}$,
respectively. Concerning the first secondary variable node, the set
$A_{s_{1}^{(p,s)}/v_{1}^{(p)}}$, of adjacent variable nodes, of the
first constituent SA check node, $s_{6}$, excluding the first primary
variable node, $v_{3}$, includes both $v_{8}$ and $v_{23}$, which
belong to users, $U_{2}$, and $U_{4}$, respectively. This means
that there are two potential candidates for the first secondary variable
node, $v_{1}^{(s)}$. Similarly, the set $A_{s_{2}^{(p,s)}/v_{2}^{(p)}}$,
of adjacent variable nodes, of the second constituent SA check node,
$s_{8}$, excluding the second primary variable node, $v_{5}$, includes
$v_{12}$, $v_{16}$, and $v_{22}$, which belong to users, $U_{2}$,
$U_{3}$, and $U_{4}$, respectively. This means that there are three
potential candidates for the second secondary variable node, $v_{2}^{(s)}$.
As a result, the algorithm should include a new nested loop, that
investigates all the possible candidate pairs of first and second
secondary variable nodes.

The example tackles three possible scenarios that could occur during
the nested loop investigation. The first scenario is concerned with
$v_{16}$, as one of the potential candidates of $v_{2}^{(s)}$. Due
to the fact that $v_{16}$ belongs to user $U_{3}$, and none of the
two potential candidates of $v_{1}^{(s)}$, belongs to $U_{3}$, no
possible pair that includes $v_{16}$, could close a global $8-$cycle.
Consequently, $v_{16}$ should be excluded. 

The second scenario investigates the pair of $v_{8}$ and $v_{12}$,
such that $v_{8}$ represents the first potential secondary variable
node, $v_{1}^{(s)}$, while $v_{12}$ represents the second potential
secondary variable node, $v_{2}^{(s)}$. Because both $v_{8}$ and
$v_{12}$, belong to the same user, $U_{2}$, they could close a global
$8-$cycle, provided that they are connected via one or more LDPC
check nodes. Analyzing the underlying parity check matrix, as described
previously, it can be shown that there is no common LDPC check node,
that connects $v_{8}$ and $v_{12}$. As a result, the least cycle
length that can be constituted by including both $v_{8}$ and $v_{12}$,
is certainly greater than eight. 

As for the third scenario, the pair of $v_{23}$ and $v_{22}$, such
that $v_{23}$ represents the first potential secondary variable node,
$v_{1}^{(s)}$, while $v_{22}$ represents the second potential secondary
variable node, $v_{2}^{(s)}$, is to be investigated. Similar to the
previous scenario, because both $v_{23}$ and $v_{22}$, belong to
the same user, $U_{4}$, they could close a global $8-$cycle, provided
that they are connected via one or more LDPC check nodes. Analyzing
the underlying parity check matrix, this time shows that $v_{23}$
and $v_{22}$, are connected via LDPC check node, $l_{20}$. Consequently,
the pair of $v_{23}$ and $v_{22}$, closes a global $8-$cycle, that
passes through $v_{3}$ and $v_{5}$, and leads to incrementing the
$8-$cycle counter, $C_{8}^{(v_{3})}$, of variable node, $v_{3}$,
by one.

\section{Results}

In the previous section, a graphical algorithm that computes the number
of global $8-$cycles, in the three-layer Tanner graph of a SCRAM
system was proposed. The proposed algorithm is pillared on the thoroughly
investigated structure of the global $8-$cycle that was presented
in section \ref{sec:Girth}. This section extends the results presented
in Section \ref{subsec:Cycle-Profile-of-SCRAM}, to include the algorithmic
calculation of the number of global $8-$cycles, in a SCRAM system,
with $N_{u}=4$ users, $N_{s}=8640$ slots, and the $(4320,2160)$
DVB-NGH \cite{gomez2014dvb} LDPC code. The aim of this section is
to analyze and validate the results obtained from the proposed algorithm,
by comparing them to the total number of $8-$cycles, from the cycle
profile of SCRAM.

Table \ref{Global 8-cycle counter} is thus regarded as an extension
to the results in Table \ref{Cycle Profile SCRAM}, with an extra
column that includes the number of global $8-$cycles, in the SCRAM
three-layer Tanner graph, obtained from the proposed global $8-$cycle
counting algorithm. Here again, the first row depicts the cycle profile
of the $(4320,2160)$ DVB-NGH \cite{gomez2014dvb} LDPC code. The
second row shows the number of global $8-$cycles, in the SCRAM three-layer
Tanner graph, obtained from the proposed global $8-$cycle counting
algorithm. In addition, it shows the cycle profile of the SCRAM system,
obtained from the Half-Cycle \cite{li2015improved} counting algorithm
of the SCRAM hybrid matrix.

Because The investigated SCRAM system incorporates $N_{u}=4$ users,
it has a number of local LDPC $8-$cycles, that corresponds to four
times the number of $8-$cycles, of a single LDPC code. This means
that, the SCRAM Tanner graph has a total of 6233360 local LDPC $8-$cycles.
Meanwhile, as shown in the table, the results of the proposed global
$8-$cycle counting algorithm, indicate that the random access scheme,
adds additional 725 global $8-$cycles. This means that the SCRAM
Tanner graph, has a total of 6234085, cycles of length eight, which
complies with the number of $8-$cycles, obtained from the cycle profile
of the Half-Cycle counting algorithm of the mapped SCRAM hybrid matrix.

The results herein show that the number of $6-$cycles in the SCRAM
three-layer Tanner graph, corresponds to four times that of the NGH
LDPC code. This indicates that all the $6-$cycles, in the SCRAM system,
are local LDPC cycles. Consequently, the hypothesis that the adopted
channel scheme does not add cycles of length six, and the claim of
Section \ref{sec:Girth}, that the shortest global cycle length is
eight, have been verified. Moreover, the results of the $8-$cycles
verify the validity of the proposed global $8-$cycle counting algorithm
of the SCRAM three-layer Tanner graph. Because the proposed algorithm
has significantly lower complexity than the Half-Cycle counting algorithm,
it can be utilized to compute the number of global $8-$cycles, of
the SCRAM system.

\begin{table}[b]
\caption{Algorithmic 8\textminus cycle results of SCRAM, with $N_{u}=4$ users,
adopting the $(4320,2160)$ DVB-NGH LDPC code, with Random Access,
on a channel with $N_{s}=8640$ slots}

\label{Global 8-cycle counter}%
\begin{tabular}{|c|c|c|c|}
\hline 
 & $C_{6}$ & $C_{8}$ & Global 8-Cycles\tabularnewline
\hline 
\hline 
$(4320,2160)$ DVB-NGH LDPC & 31200 & 1558340 & \tabularnewline
\hline 
SCRAM, with $N_{u}=4$ users & 124800 & 6234085 & 725\tabularnewline
\hline 
\end{tabular}
\end{table}

\section{Conclusions}

In this paper, a graphical correlation-based method that counts the
number of global $8-$cycles on the SCRAM (Slotted Coded Random Access
Multiplexing) three-layer Tanner graph is proposed. The essence of
SCRAM lies in its hybrid decoding structure, that jointly resolves
the collisions and decodes the Low Density Parity Check (LDPC) codewords,
in a similar analogy to Belief Propagation Decoding, on a joint three-layer
Tanner graph. This paper leverages the analogy between the two-layer
Tanner graph of classical LDPC codes, and the three-layer Tanner graph
of the proposed SCRAM, in order to further optimize the performance
of SCRAM. 

The paper first introduces the necessary procedure required to utilize
the leading state-of-the-art tools that compute the cycle profile
of classical LDPC codes, to compute the cycle profile of SCRAM. The
paper also derives a lower bound on the shortest cycle length (girth)
of a SCRAM system, with arbitrary number of accommodated users and
deployed LDPC codes. In essence, the paper proposes a novel graphical
approach that scans the three-layer Tanner graph of SCRAM, in search
for the cycles of girth length. The results presented herein show
that the proposed cycle-counting approach yields exactly the same
results as the conventional LDPC tools at much lower complexity cost.
Future implications include the utilization of the analysis tools
proposed herein, in order to devise an optimized SCRAM system with
bounds on the cycle profile as the optimization cost function.

\appendices{}

\bibliographystyle{IEEEtran}
\bibliography{literature}

\begin{thebibliography}{10}
\providecommand{\url}[1]{#1}
\csname url@samestyle\endcsname
\providecommand{\newblock}{\relax}
\providecommand{\bibinfo}[2]{#2}
\providecommand{\BIBentrySTDinterwordspacing}{\spaceskip=0pt\relax}
\providecommand{\BIBentryALTinterwordstretchfactor}{4}
\providecommand{\BIBentryALTinterwordspacing}{\spaceskip=\fontdimen2\font plus
\BIBentryALTinterwordstretchfactor\fontdimen3\font minus
  \fontdimen4\font\relax}
\providecommand{\BIBforeignlanguage}[2]{{%
\expandafter\ifx\csname l@#1\endcsname\relax
\typeout{** WARNING: IEEEtran.bst: No hyphenation pattern has been}%
\typeout{** loaded for the language `#1'. Using the pattern for}%
\typeout{** the default language instead.}%
\else
\language=\csname l@#1\endcsname
\fi
#2}}
\providecommand{\BIBdecl}{\relax}
\BIBdecl

\bibitem{miraz2015review}
M.~H. Miraz, M.~Ali, P.~S. Excell, and R.~Picking, ``A review on internet of
  things ({I}o{T}), internet of everything ({I}o{E}) and internet of nano
  things ({I}o{NT}),'' \emph{2015 Internet Technologies and Applications
  (ITA)}, pp. 219--224, 2015.

\bibitem{saad2019vision}
W.~Saad, M.~Bennis, and M.~Chen, ``A vision of {6G} wireless systems:
  Applications, trends, technologies, and open research problems,'' \emph{IEEE
  network}, vol.~34, no.~3, pp. 134--142, 2019.

\bibitem{chowdhury20206g}
M.~Z. Chowdhury, M.~Shahjalal, S.~Ahmed, and Y.~M. Jang, ``{6G} wireless
  communication systems: Applications, requirements, technologies, challenges,
  and research directions,'' \emph{IEEE Open Journal of the Communications
  Society}, vol.~1, pp. 957--975, 2020.

\bibitem{ding2015cooperative}
Z.~Ding, M.~Peng, and H.~V. Poor, ``Cooperative non-orthogonal multiple access
  in {5G} systems,'' \emph{IEEE Communications Letters}, vol.~19, no.~8, pp.
  1462--1465, 2015.

\bibitem{islam2016power}
S.~R. Islam, N.~Avazov, O.~A. Dobre, and K.-S. Kwak, ``Power-domain
  non-orthogonal multiple access {(NOMA)} in {5G} systems: Potentials and
  challenges,'' \emph{IEEE Communications Surveys \& Tutorials}, vol.~19,
  no.~2, pp. 721--742, 2016.

\bibitem{reddy2021analytical}
P.~V. Reddy, S.~Reddy, S.~Reddy, R.~D. Sawale, P.~Narendar, C.~Duggineni, and
  H.~B. Valiveti, ``Analytical review on {OMA} vs. {NOMA} and challenges
  implementing {NOMA},'' in \emph{2021 2nd International Conference on Smart
  Electronics and Communication (ICOSEC)}.\hskip 1em plus 0.5em minus
  0.4em\relax IEEE, 2021, pp. 552--556.

\bibitem{srivastava2021non}
S.~Srivastava and P.~P. Dash, ``Non-orthogonal multiple access: Procession
  towards {B5G} and {6G},'' in \emph{2021 IEEE 2nd International Conference on
  Applied Electromagnetics, Signal Processing, \& Communication (AESPC)}.\hskip
  1em plus 0.5em minus 0.4em\relax IEEE, 2021, pp. 1--4.

\bibitem{maraqa2020survey}
O.~Maraqa, A.~S. Rajasekaran, S.~Al-Ahmadi, H.~Yanikomeroglu, and S.~M. Sait,
  ``A survey of rate-optimal power domain {NOMA} with enabling technologies of
  future wireless networks,'' \emph{IEEE Communications Surveys \& Tutorials},
  vol.~22, no.~4, pp. 2192--2235, 2020.

\bibitem{lee2020user}
I.-H. Lee and H.~Jung, ``User selection and power allocation for downlink
  {NOMA} systems with quality-based feedback in rayleigh fading channels,''
  \emph{IEEE Wireless Communications Letters}, vol.~9, no.~11, pp. 1924--1927,
  2020.

\bibitem{gupta2020user}
P.~Gupta and D.~Ghosh, ``User fairness based energy efficient power allocation
  for downlink cellular {NOMA} system,'' in \emph{2020 5th International
  Conference on Computing, Communication and Security (ICCCS)}.\hskip 1em plus
  0.5em minus 0.4em\relax IEEE, 2020, pp. 1--5.

\bibitem{jehan2022comparative}
A.~Jehan and M.~Zeeshan, ``Comparative performance analysis of code-domain
  {NOMA} and power-domain {NOMA},'' in \emph{2022 16th International Conference
  on Ubiquitous Information Management and Communication (IMCOM)}.\hskip 1em
  plus 0.5em minus 0.4em\relax IEEE, 2022, pp. 1--6.

\bibitem{liu2021sparse}
Z.~Liu and L.-L. Yang, ``Sparse or dense: A comparative study of code-domain
  {NOMA} systems,'' \emph{IEEE Transactions on Wireless Communications},
  vol.~20, no.~8, pp. 4768--4780, 2021.

\bibitem{chen2016pattern}
S.~Chen, B.~Ren, Q.~Gao, S.~Kang, S.~Sun, and K.~Niu, ``Pattern division
  multiple access-a novel nonorthogonal multiple access for fifth-generation
  radio networks,'' \emph{IEEE Transactions on Vehicular Technology}, vol.~66,
  no.~4, pp. 3185--3196, 2016.

\bibitem{nikopour2013sparse}
H.~Nikopour and H.~Baligh, ``Sparse code multiple access,'' in \emph{2013 IEEE
  24th Annual International Symposium on Personal, Indoor, and Mobile Radio
  Communications (PIMRC)}.\hskip 1em plus 0.5em minus 0.4em\relax IEEE, 2013,
  pp. 332--336.

\bibitem{al2014uplink}
M.~Al-Imari, P.~Xiao, M.~A. Imran, and R.~Tafazolli, ``Uplink non-orthogonal
  multiple access for {5G} wireless networks,'' in \emph{2014 11th
  international symposium on wireless communications systems (ISWCS)}.\hskip
  1em plus 0.5em minus 0.4em\relax IEEE, 2014, pp. 781--785.

\bibitem{chaturvedi2022tutorial}
S.~Chaturvedi, Z.~Liu, V.~A. Bohara, A.~Srivastava, and P.~Xiao, ``A tutorial
  on decoding techniques of sparse code multiple access,'' \emph{IEEE Access},
  2022.

\bibitem{zhang2022hybrid}
N.~Zhang and X.~Zhu, ``A hybrid grant {NOMA} random access for massive {MTC}
  service,'' \emph{IEEE Internet of Things Journal}, vol.~10, no.~6, pp.
  5490--5505, 2022.

\bibitem{gollakota2008zigzag}
S.~Gollakota and D.~Katabi, ``Zigzag decoding: Combating hidden terminals in
  wireless networks,'' in \emph{Proceedings of the ACM SIGCOMM 2008 conference
  on Data communication}, 2008, pp. 159--170.

\bibitem{tehrani2011sigsag}
A.~S. Tehrani, A.~G. Dimakis, and M.~J. Neely, ``Sigsag: Iterative detection
  through soft message-passing,'' \emph{IEEE Journal of Selected Topics in
  Signal Processing}, vol.~5, no.~8, pp. 1512--1523, 2011.

\bibitem{xiao2019low}
J.~Xiao, J.~Hu, and K.~Han, ``Low complexity expectation propagation detection
  for scma using approximate computing,'' in \emph{2019 IEEE Global
  Communications Conference (GLOBECOM)}.\hskip 1em plus 0.5em minus 0.4em\relax
  IEEE, 2019, pp. 1--6.

\bibitem{herath2020low}
P.~Herath, A.~Haghighat, and L.~Canonne-Velasquez, ``A low-complexity
  interference cancellation approach for {NOMA},'' in \emph{2020 IEEE 91st
  Vehicular Technology Conference (VTC2020-Spring)}.\hskip 1em plus 0.5em minus
  0.4em\relax IEEE, 2020, pp. 1--5.

\bibitem{iswarya2021survey}
N.~Iswarya and L.~Jayashree, ``A survey on successive interference cancellation
  schemes in non-orthogonal multiple access for future radio access,''
  \emph{Wireless Personal Communications}, vol. 120, no.~2, pp. 1057--1078,
  2021.

\bibitem{ling2017multiple}
B.~Ling, C.~Dong, J.~Dai, and J.~Lin, ``Multiple decision aided successive
  interference cancellation receiver for {NOMA} systems,'' \emph{IEEE Wireless
  Communications Letters}, vol.~6, no.~4, pp. 498--501, 2017.

\bibitem{ren2016advanced}
B.~Ren, X.~Yue, W.~Tang, Y.~Wang, S.~Kang, X.~Dai, and S.~Sun, ``Advanced {IDD}
  receiver for {PDMA} uplink system,'' in \emph{2016 IEEE/CIC International
  Conference on Communications in China (ICCC)}.\hskip 1em plus 0.5em minus
  0.4em\relax IEEE, 2016, pp. 1--6.

\bibitem{nafie2018scram}
S.~Nafie, J.~Robert, and A.~Heuberger, ``{SCRAM}: A novel approach for reliable
  ultra-low latency m2m applications,'' in \emph{2018 IEEE 88th Vehicular
  Technology Conference (VTC-Fall)}.\hskip 1em plus 0.5em minus 0.4em\relax
  IEEE, 2018, pp. 1--5.

\bibitem{munari2015multi}
A.~Munari, F.~Clazzer, and G.~Liva, ``Multi-receiver {ALOHA} systems-a survey
  and new results,'' in \emph{Communication Workshop (ICCW), 2015 IEEE
  International Conference on}.\hskip 1em plus 0.5em minus 0.4em\relax IEEE,
  2015, pp. 2108--2114.

\bibitem{gallager1962low}
R.~Gallager, ``Low-density parity-check codes,'' \emph{IRE Transactions on
  information theory}, vol.~8, no.~1, pp. 21--28, 1962.

\bibitem{mackay1997near}
D.~J. MacKay and R.~M. Neal, ``Near shannon limit performance of low density
  parity check codes,'' \emph{Electronics letters}, vol.~33, no.~6, pp.
  457--458, 1997.

\bibitem{tanner1981recursive}
R.~Tanner, ``A recursive approach to low complexity codes,'' \emph{IEEE
  Transactions on information theory}, vol.~27, no.~5, pp. 533--547, 1981.

\bibitem{karimi2012message}
M.~Karimi and A.~H. Banihashemi, ``Message-passing algorithms for counting
  short cycles in a graph,'' \emph{IEEE Transactions on Communications},
  vol.~61, no.~2, pp. 485--495, 2012.

\bibitem{li2015improved}
J.~Li, S.~Lin, and K.~Abdel-Ghaffar, ``Improved message-passing algorithm for
  counting short cycles in bipartite graphs,'' in \emph{2015 IEEE International
  Symposium on Information Theory (ISIT)}.\hskip 1em plus 0.5em minus
  0.4em\relax IEEE, 2015, pp. 416--420.

\bibitem{johnson2006introducing}
S.~J. Johnson, ``Introducing low-density parity-check codes,'' \emph{University
  of Newcastle, Australia}, vol.~1, p. 2006, 2006.

\bibitem{xiao2019reed}
X.~Xiao, B.~Vasi{\'c}, S.~Lin, K.~Abdel-Ghaffar, and W.~E. Ryan, ``Reed-solomon
  based quasi-cyclic {LDPC} codes: Designs, girth, cycle structure, and
  reduction of short cycles,'' \emph{IEEE Transactions on Communications},
  vol.~67, no.~8, pp. 5275--5286, 2019.

\bibitem{gomez2014dvb}
D.~Gomez-Barquero, C.~Douillard, P.~Moss, and V.~Mignone, ``{DVB-NGH}: The next
  generation of digital broadcast services to handheld devices,'' \emph{IEEE
  Transactions on Broadcasting}, vol.~60, no.~2, pp. 246--257, 2014.

\end{thebibliography}

\end{document}